\documentclass[extra,referee]{gji}
\usepackage{times}
\usepackage{amsmath,amsfonts,bm,graphicx,color}
\usepackage[normalem]{ulem}
\newcommand{\argmin}{\mathop{\rm argmin}}

\title[Geophys.\ J.\ Int.: FWI with NL similarity and model-derivative domain dictionaries]
  {Full waveform inversion with nonlocal similarity and model-derivative domain adaptive sparsity-promoting regularization}
\author[D. Li \& J. Harris]
  {Dongzhuo Li and Jerry M. Harris\\
  Department of Geophysics, Stanford University, Stanford, CA \emph{94305}, USA.
   Email: lidongzh@stanford.edu}
\date{}
\pagerange{\pageref{firstpage}--\pageref{laspage}}
\volume{XXX}
\pubyear{2018}

\begin{document}

\label{firstpage}

\maketitle

\begin{summary}
Full waveform inversion (FWI) is a highly nonlinear and ill-posed problem. On one hand, it can be easily trapped in a local minimum. On the other hand, the inversion results may exhibit strong artifacts and reduced resolution because of inadequate constraint from data. Proper regularizations are necessary to reduce such artifacts and steer the inversion towards a good direction.


In this study, we propose a novel adaptive sparsity-promoting regularization for FWI in the model-derivative domain which exploits nonlocal similarity in the model. This regularization can be viewed as a generalization of total variation (TV) with multi-class learning-based dictionaries. The dictionaries incorporate the prior information of nonlocal similarity into the inversion, exploiting the fact that geological patterns at different places are similar to some others up to affine transformations (translation, rotation and scaling). Such nonlocal similarity priors effectively reduce the degrees of freedom in model parameters, and may also mitigate the problem of local minima. The formulated optimization problem is solved by the Alternating Direction Method of Multipliers (ADMM). By interpreting the iterative scheme, we find our method closely connected with image processing techniques and convolutional neural networks (CNN).

We test our proposed method on a modified BP 2004 velocity model and a smoothed Marmousi model. Compared with traditional FWI, our technique can better reconstruct sharp edges such as salt body boundaries, and is also able to effectively reduce artifacts. Compared with TV, our result is less blocky and more geologically realistic. Quantitatively, our result has the highest structural similarity index (SSIM) and also the lowest model mean square error. This proposed method provides a general framework for incorporating nonlocal similarity priors into inversion, and can be applied to other (transformed) domains with the double-sparsity strategy.


\end{summary}


\section{Introduction}
Full waveform inversion (FWI) is one of the state-of-the-art techniques for subsurface model parameter estimation. By minimizing the difference between the observed and synthetic data, FWI seeks to find a model that explains the observed data as well as possible. However, to solve this wave-equation-constrained nonlinear inverse problem, we are faced with at least two intrinsic difficulties.

First, in a typical seismic survey, the aperture size and source frequency bandwidth are limited. Also, some parts of the model, such as sub-salt regions, may not be well illuminated due to the complicated wave phenomena in heterogeneous media. For these reasons, model parameters are not well constrained by the data. Therefore, on one hand, the wavenumber spectrum of geological structures cannot be fully recovered, resulting in reduced resolution \citep{virieux2009overview}. On the other hand, strong artifacts may be generated due to this lack of constraint.

Second, as a nonlinear inverse problem that is almost exclusively solved by local optimization methods, FWI can easily fall into a local minimum if the initial model is not sufficiently accurate and the objective function is highly non-convex. For example, the $L_2$ norm commonly used in data misfit calculations is highly nonlinear with respect to time shifts in data. This may cause the widely-known ``cycle-skipping'' problem \citep{warner2016adaptive,metivier2016measuring}.

To address these problems, one way is to add regularization into FWI. First, although we can have millions or billions of model parameters, there are actually not so many degrees of freedom. For example, we know that the model has certain degrees of smoothness or blockiness, and geological structures should have mutual coherency. In other words, the possible models actually reside on a low-dimensional space embedded in a high-dimensional space. Regularization may reduce the aforementioned artifacts and enhance resolution by guiding the inversion towards such a low-dimensional space, with predefined prior knowledge complementary to the input data. Second, regularization can change or ``convexify'' the misfit surface, or restrict possible solutions to regions where the objective function is more convex, which may mildly mitigate the problem of local minima.


As a widely-used regularization method, Tikhonov regularization is simple to implement and effective in stabilizing inversion by recovering the minimum-norm solution. However, it promotes smoothed inversion results and is unable to recover some small-scale features \citep{bovik2010handbook,Aster201393}. 

Another category of regularization methods assumes that the model is sparse by itself or sparse in a transform domain, and tries to find the sparsest model in that domain through inversion. This sparsity-promoting regularization has the potential to recover high-resolution structures. Examples include tomography with sparsity constraints in the wavelet domain \citep{loris2007tomographic,simons2011solving,charlety2013global}, sparsity promoting FWI in the curvelet domain \citep{li2012fast} and in the seislet domain \citep{xue2017full}. One important regularization in this category is Total Variation regularization (TV).  It was first proposed for image denoising \citep{rudin1992nonlinear} and has been successfully applied to other image processing tasks, such as debluring \citep{chan1998total} and super-resolution \citep{marquina2008image}. TV regularization requires that the model be sparse in the model-derivative domain, and thus favors piecewise smooth structures while preserves edges, which are geological features of most sedimentary layers and intrusive salt bodies. Therefore, FWI with TV regularization and other variants of blocky regularizations \citep{maharramov2014robust,guitton2012blocky,esser2016total} are used to make use of this prior. TV regularization is also shown to be useful in geothermal reservoir delineation \citep{lin2014acoustic}, and CO2 injection monitoring \citep{zhu2015applications}. In this paper, we use model derivatives, which are often called gradients in the imaging processing community, to denote the vertical or horizontal first-order finite differences of the model. When we are not discussing gradients in optimization problems, we use model derivatives and model gradients interchangeably.

The transform methods/domains described above are analytical, and cannot adapt to patterns in specific problems. Hence, learning-based dictionaries such as overcomplete dictionaries from the KSVD algorithm \citep{aharon2006img} and orthogonal dictionaries \citep{bao2013fast} have been proposed for sparsity-promoting problems. By learning from a training set of patterns, such dictionaries can be finely tuned to have sparser representations of patterns than analytic dictionaries. \cite{zhu2017sparse} used online orthogonal dictionaries in FWI to reduce the artifacts in inversion with supershots.

Nonlocal similarity posits that different parts of model may be geometrically similar. It was first introduced by the image processing community and has been applied to image denoising \citep{buades2005review,dabov2007image}, image restoration and super-resolution \citep{dong2011image,glasner2009super,huang2015single}. In such inversion tasks, nonlocal similarity can serve as a powerful prior to effectively reduce the degrees of freedom. We know of little work applying nonlocal similarity to full waveform inversion. The closest is a nonlocal total variation (NLTV) regularization technique \citep{gilboa2008nonlocal} which uses nonlocal differential operators, but it is different from our approach, which can be generalized to other domains rather than the model-derivative domain only.

In this research, we propose a novel full waveform inversion method with adaptive sparsity-promoting regularization in the model-derivative domain that exploits nonlocal similarity in the model. Specifically, we introduce a nonlocal similarity prior into FWI. The reason why we work in the model-derivative domain is that in this domain features are mainly edges or boundaries, so that it is easier to find similar ones across the model (Figure \ref{fig:nl_similarity}). Inspirations for this work, in addition to what has been discussed above, also come from the double-sparsity dictionary model \citep{liu2013adaptive} and multi-class dictionary learning \citep{son2014local} from the image processing community. Our method can be viewed as a generalization of TV regularization with an added layer of adaptiveness introduced by learning-based dictionaries. In our method, the dictionaries and the multi-class strategy allow us to share information between different parts of the model based on their nonlocal similarity. We regularize the structures of the model derivatives by approximating them using linear combinations of only a few important learned patterns, thereby reducing noise and artifacts in the model derivatives. All these benefits can help to steer the inversion towards a good direction.


In the following sections, we first review the background of acoustic FWI and dictionary learning. Then, we introduce the theory of our new method and corresponding numerical algorithm. Finally, we apply the new method to a modified BP 2004 model and a smoothed Marmousi model, and evaluate its performance both qualitatively and quantitatively. We also conduct a series of sensitivity tests to measure the effects of parameter selection on the inversion results.


\section{Background Theory}
\subsection{Acoustic Full Waveform Inversion}
In this section, we briefly introduce the principles of traditional acoustic full waveform inversion that our new method builds up on. 

In a seismic survey, the source and receiver arrays are deployed according to some designed survey geometry, such as surface seismic, vertical seismic profiles, or cross-well experiments. At each receiver location $\mathbf{x}_r$, we can record the observed waveform $d_{\text{obs}}(\mathbf{x}_r, t; \mathbf{x}_s)$ generated from source location $\mathbf{x}_s$. The wave propagation process is traditionally approximated by the acoustic wave equation as in \citet{Nolet2008}: 
\begin{equation}
\label{eq:AcousticWaveEq}
	\left[\frac{1}{c^2(\mathbf{x})}\frac{\partial^2}{\partial t^2}  - \rho(\mathbf{x}) \nabla \cdot \left(\frac{1}{\rho(\mathbf{x})}\nabla\right) \right] p(\mathbf{x}, t) = s(\mathbf{x}, t),
\end{equation}
where $p(\mathbf{x})$ is the pressure wavefield generated by source $s(\mathbf{x}, t)$ in the given medium parameterized by velocity $c(\mathbf{x})$ and density $\rho(\mathbf{x})$. With the acoustic wave equation, we can numerically compute the calculated waveform $d_{\text{cal}}(\mathbf{x}_r, t; \mathbf{x}_s; \mathbf{m})$, where $\mathbf{m}$ stands for the model parameters to be inverted for, that is, we have $\mathbf{m(x)} = \left[c(\mathbf{x}) \, \rho(\mathbf{x}) \, s(\mathbf{x})\right]^T$ in the most general case. In this study, however, we ignore density constrasts and only invert for velocity, and thus $\mathbf{m(x)} = c(\mathbf{x})$. Equation (\ref{eq:AcousticWaveEq}) becomes the constant-density acoustic wave equation:
\begin{equation}
\label{eq:ConstantAcousticWaveEq}
	\left[\frac{1}{c^2(\mathbf{x})}\frac{\partial^2}{\partial t^2}  - \nabla^2 \right] p(\mathbf{x}, t) = s(\mathbf{x}, t).
\end{equation}

Now we would like to minimize the difference between the observed and calculated waveforms in order to come up with a model $\mathbf{m}$ that can explain the observed data. The difference between $d_{\text{obs}}$ and $d_{\text{cal}}$ is quantified by a misfit function, such as the $L_2$-norm of their difference:
\begin{equation}
\label{eq:data_misfit}
\begin{split}
	\chi &= \frac{1}{2}\sum_{s,r} \left\Vert d_{\text{obs}} \left( \mathbf{x}_r, t; \mathbf{x}_s \right) - d_{\text{cal}} \bigl( \mathbf{x}_r, t; \mathbf{x}_s; c \left(\mathbf{x} \right) \bigr) \right \Vert_2^2 \\
	&= \frac{1}{2}\sum_{s,r}\int^T_0\left\vert(d_{\text{obs}}\left(\mathbf{x}_r, t; \mathbf{x}_s\right) - d_{\text{cal}}\bigl(\mathbf{x}_r, t; \mathbf{x}_s; c\left(\mathbf{x}\right)\bigr)\right\vert^2 \mathrm{d}t.
\end{split}
\end{equation}
Other measures, such as cross-correlation traveltime \citep{luo1991wave} can also be used, and may be more robust when dealing with legacy field data, whose amplitudes were not faithfully recorded.

To minimize $\chi$, we take its variation and establish a linear relationship between velocity perturbation $\delta c(\mathbf{x})$ and the misfit function perturbation $\delta \chi$ as follows:
\begin{equation}
\label{eq:chi_variations}
	\delta \chi = \int_V \delta c(\mathbf{x}) K(\mathbf{x}) \mathrm{d} \mathbf{x},
\end{equation}
where according to the adjoint method proposed by \cite{Tarantola-147}, the Fr\'echet kernel can be derived as
\begin{equation}
	K(\mathbf{x}) = \frac{2}{c_0^3(\mathbf{x})} \sum_{s} \int^T_0 \dot{p}(\mathbf{x}, t; \mathbf{x}_s) \, \dot{p}^{\dagger}(\mathbf{x},T-t;\mathbf{x}_s) \mathrm{d}t,
\end{equation}
where $c_0(\mathbf{x})$ is the background velocity field, and $\dot{p}^{\dagger}(\mathbf{x},T-t;\mathbf{x}_s)$ is the time derivative of the adjoint wavefield computed by back propagating the data residuals. 

In the discrete case, the acoustic velocity field is parameterized by basis functions $b_i(\mathbf{x})$ such as the natural basis (pixels), splines or spherical harmonics. Formally, $c(\mathbf{x}) = \sum_i c_i b_i(\mathbf{x})$, and hence $\delta c(\mathbf{x}) = \sum_i \delta c_i b_i(\mathbf{x})$. Thus Equation (\ref{eq:chi_variations}) becomes
\begin{equation}
	\delta \chi = \sum_i \delta c_i \int_V K(\mathbf{x}) b_i(\mathbf{x}) \mathrm{d} \mathbf{x},
\end{equation}
and hence we obtain the gradient of misfit function $\chi$ with respect to the velocity coefficient vector $\mathbf{c}$:
\begin{equation}
	\nabla_{\mathbf{c}}\chi = \left[\int_{V} K(\mathbf{x}) b_1(\mathbf{x}) \mathrm{d} \mathbf{x}, \int_{V} K(\mathbf{x}) b_2(\mathbf{x}) \mathrm{d} \mathbf{x}, \cdots, \int_{V} K(\mathbf{x}) b_n(\mathbf{x}) \mathrm{d} \mathbf{x}\right]^T,
\end{equation}
which is just the projection of the kernel onto the basis functions \citep{fichtner2010full}.

Once we have the gradient information, we can employ any gradient-based optimization method, such as the steepest-descent method, conjugate gradient method, or quasi-Newton method, to solve this problem. In this study, we adopt the L-BFGS method \citep{nocedal2006numerical}, a type of quasi-Newton method, due to its fast convergence speed and the ability to compensate limited illumination by computing an approximate Hessian \citep{pratt1998gauss}.

\subsection{Dictionary Learning}
\label{sec:dl}
A dictionary $\mathbf{D}$ is a matrix of dimension ${n_d\times n_a}$, which consists of $n_a$ column vectors or atoms of dimension $n_d$. A dictionary is either complete or overcomplete, so any signal of the same dimension as the atoms can be expressed by linear combinations of atoms in the dictionary. Usually, dictionaries are used for sparse approximation of the given signals. Sparsity means that only a few entries in a signal are nonzero, or that $\#\{i: x_i \neq 0\}$ is small.

The sparse approximation problem has the following form
\begin{equation}
	\begin{array}{ll}
		\mbox{minimize} & \|\mathbf{x}\|_0 \\
		\mbox{subject to} & \|\mathbf{y} - \mathbf{Dx}\|_2 \leq \epsilon,
	\end{array}
\end{equation}
where $\mathbf{y}$ is the original signal, $\mathbf{D}$ the dictionary, $\mathbf{x}$ the sparse representation coefficients, $\epsilon$ the approximation error tolerance, and $\|\cdot\|_0$ stands for the $\ell_0$ quasi-norm. This problem finds the sparsest representation of $\mathbf{y}$ under the constraint that the approximation error is below a given tolerance level $\epsilon$. 

The reasons for using sparse approximation are two-fold. First, sparsity is a good prior knowledge as natural signals such as images are sparse under the representation of certain dictionaries. Second, by keeping only a few important elements we can reduce noise and artifacts and meanwhile preserve small-scale geological structures.

There are two types of dictionaries, analytic and learning-based. Examples of analytic dictionaries include wavelets \citep{daubechies1988orthonormal,daubechies2004iterative}, curvelets \citep{herrmann2008non,herrmann2007curvelet}, seislets \citep{fomel2010seislet,dutta2015sparse,yang2018time} and so on. They are designed to deal with a wide range of signals and therefore might not give the sparsest representation for a given type of signals. On the contrary, learning-based dictionaries are generated by training, so they are more adapted to the specific signal patterns in a given problem, yielding sparser representations of the signals, and are thus better at denoising and preserving structures.




Mathematically, one way to write the objective function for sparse dictionary learning is
\begin{equation}
\label{eq:dictLearn}
	\argmin_{\mathbf{D}, \mathbf{X}} \|\mathbf{Y} - \mathbf{DX}\|_F^2, \, \text{s.t. }\|\mathbf{x}_i\|_0 < T_0, \forall i,
\end{equation}
where matrix $\mathbf{Y}$ contains signal vectors $\{\mathbf{y}_i\}_{i=1}^{n_a}$, matrix $\mathbf{X}$ contains the corresponding coefficient vectors $\{\mathbf{x}_i\}_{i=1}^{n_a}$, $F$ denotes the Frobenius norm, and $T_0$ controls the number of non-zero entries in each column vector $\mathbf{x}_i$. Problem (\ref{eq:dictLearn}) can be solved by the method proposed by \cite{lee2007efficient} or the K-SVD algorithm \citep{aharon2006img,lee2007efficient}. The dictionaries generated are usually overcomplete, which means that the number of atoms in a dictionary can be larger than the dimension of the atom signals ($n_d < n_a$). The atoms in an overcomplete dictionary are not linear independent, let alone orthogonal to each other. A learning-based overcomplete dictionaries may offer better sparse approximation properties than analytic bases or frames \citep{rubinstein2010dictionaries}. However, overcomplete dictionary learning has much higher computational complexity.

Another type of dictionary learning method is the orthogonal dictionary learning \citep{bao2013fast}. In the formulation of \cite{bao2013fast}, the inequality constraint in Problem (\ref{eq:dictLearn}) is absorbed into the objective function by additionally minimizing the $\ell_0$-norm of the coefficient vectors $\mathbf{x}_i$. Herein, we change Bao's original $\ell_0$-norm to $\ell_1$-norm, which also promotes sparsity and is used in TV regularization, so this helps to ensure a fair comparison between our proposed method and TV. The dictionary learning problem has one orthogonality constraint:
\begin{equation}
\label{eq:orthoDictLearn}
	\argmin_{\mathbf{D, X}} \|\mathbf{Y} - \mathbf{DX}\|_F^2 + \lambda \sum_l\|\mathbf{X}(:,l)\|_1, \, \text{s.t. } \mathbf{D}^T \mathbf{D} = \mathbf{I},
\end{equation}
where $\mathbf{D, X, Y}$ are defined the same as before, $\lambda$ is a parameter that controls the weight of sparsity promotion, and $\mathbf{I}$ is the idendity matrix. Compared to overcomplete dictionary learning, orthogonal dictionary learning is much faster. Also, due to orthogonality, it is faster to perform forward and backward transformations between signals and their coefficients. Besides, orthogonal dictionaries have lower mutual incoherency \citep{tropp2004greed} than overcomplete ones, and would yield better sparse approximation performances \citep{bao2013fast}. Based on these two facts, we adopt the orthogonal dictionary learning method in this study. One drawback of orthogonal dictionaries, however, is that they do not contain as many patterns as overcomplete dictionaries. We will discuss in the next section about how to get around this limitation.

Problem (\ref{eq:orthoDictLearn}) can be efficiently solved by alternating between a sparse coding stage and a dictionary updating stage. To be specific, the first subproblem is
\begin{equation}
\label{eq:dicttion_learning_X}
	\argmin_{\mathbf{X}} \|\mathbf{Y} - \mathbf{DX}\|_F^2 + \lambda \sum_l\|\mathbf{X}(:,l)\|_1,
\end{equation}
which can be solved by soft-thresholding $\mathbf{X} = S_{\lambda/2}(\mathbf{D}^T \mathbf{Y})$, where $S_\lambda$ is an element-wise soft-thresholding operator \citep{hastie2009elements}, defined as
\begin{equation}
	S_\lambda (x_i) = 
	\begin{cases}
				x_i - \lambda, & \text{if}\, x_i > \lambda\\
	      0, & \text{if}\, |x_i| \leq \lambda \\
	      x_i + \lambda , & \text{if}\, x_i < -\lambda
    	\end{cases}.
\end{equation}
The second subproblem is
\begin{equation}
\label{eq:dicttion_learning_D}
	\argmin_{\mathbf{D}} \|\mathbf{Y} - \mathbf{DX}\|_F^2, \, \text{s.t. } \mathbf{D}^T \mathbf{D} = \mathbf{I},
\end{equation}
which has the unique solution $\mathbf{D} = \mathbf{UV}^T$, where $\mathbf{U}$ and $\mathbf{V}$ are from the SVD of $\mathbf{YX}^T$: $\mathbf{YX}^T = \mathbf{U}\bm{\Sigma}\mathbf{V}^T$.

\section{Proposed Method -- NMAS-FWI}
As stated in the introduction, we need to add prior knowledge of model parameters to guide the inversion process to a reasonable solution. This is usually accomplished by adding a regularization term. The inversion problem then becomes
\begin{equation}
	\min_{\mathbf{m}} \mathcal{C}_d(\mathbf{m}) + \lambda \mathcal{C}_{\mathbf{m}}(\mathbf{m}),
\end{equation}
where $\mathcal{C}_d(\mathbf{m})$ is the data misfit term such as Equation (\ref{eq:data_misfit}), $\mathcal{C}_{\mathbf{m}}(\mathbf{m})$ is the regularization term, and the weighting parameter $\lambda$ controls the balance of importance between these two terms.

In this study, we are using sparsity-promoting regularization with multi-class dictionary learning. This algorithm works in the model-derivative domain, mainly based on the consideration that the model derivative fields are naturally sparse. Also, the major features in the model-derivative domain are edges, and it is easy to find similar features to them at different locations or at different rotation angles and scales. We call this method \textbf{N}onlocal similarity \textbf{M}odel-derivative domain \textbf{A}daptive \textbf{S}parsity-promoting algorithm, or NMAS method for short. 

In the following subsections, we push the mathematical part to the front. For those who would like to grasp the intuitions behind the equations first, we recommend that they read subsection \ref{subsec:NLS} and \ref{sec:mcd} before 3.1.

\subsection{The Objective Function \& Algorithm}
\label{sec:objfunc}
\textbf{Objective Function}

We denote the data vector by $\mathbf{d} \in \mathbb{R}^{N_s N_r N_t}$, meaning that there are $N_s$ sources, $N_r$ receivers, and in each data trace there are $N_t$ time steps. All data traces are stacked together into the data vector $\mathbf{d}$. If the discretized model is 2-D for example, and has a vertical dimension of $M$ and a horizontal dimension of $N$, we denote the model vector by $\mathbf{m} \in \mathbb{R}^{MN}$.  The forward modeling operator $f: \mathbb{R}^{MN} \rightarrow \mathbb{R}^{N_s N_r N_t}$ maps from the model space to the data space by solving the constant-density acoustic wave equation (\ref{eq:ConstantAcousticWaveEq}).

We also define a patch extraction operator $\mathbf{R}_i: \mathbb{R}^{MN} \rightarrow \mathbb{R}^{n^2}$, such that $\mathbf{R}_i \mathbf{m} = \mathbf{x}_i \in \mathbb{R}^{n^2}, \, i=1, 2, \cdots, N_p (\text{total number of patches})$, and each patch vector $\mathbf{x}_i$ corresponds to a $n\times n$ 2-D patch in the original model. Also, its adjoint operator $\mathbf{R}_i^T: \mathbb{R}^{n^2} \rightarrow \mathbb{R}^{MN}$ puts the patch vector back to the model vector. In practice, the patches are extracted by sliding a $n\times n$ window across the model in a column-major sequence. We choose the moving stride to be 1, which means that the patches have maximum overlap, and assume periodic boundary conditions (the patches at the boundaries wrap around to the opposite side). We will see later that this is beneficial since $\sum_i \mathbf{R}_i^T \mathbf{R}_i = n^2\mathbf{I}$ with that assumption, where $\mathbf{I}$ is the identity matrix.

With these definitions, we formulate the objective function of our proposed method as
\begin{equation}
\label{eq:objective}
\argmin_{\mathbf{m}} \frac{1}{2}\|\mathbf{d} - f(\mathbf{m})\|^2_2 
+ \beta \sum_{i,j}\|\mathbf{D}^T_{c(i),j} \mathbf{R}_{i}\bm{\nabla}_j \mathbf{m}\|_1,
\end{equation}
which consists of two parts: the first one is the data misfit term while the second one is the regularization term, and they are balanced by a trade-off parameter $\beta$. In the regularization term, the directional finite difference operators $\bm{\nabla}_j$, where $j = 1 \text{ or } 2$ stands for the vertical or horizontal direction respectively, are applied to the model $\mathbf{m}$ to compute the derivatives of the model. Then the patch extraction operator $\mathbf{R}_i$ extracts the $i$-th patch, and we easily find the coefficients of that patch in an orthogonal dictionary $\mathbf{D}_{c(i),j}$, using the property that the transpose of the dictionary equals its inverse: $\mathbf{D}^T_{c(i),j} = \mathbf{D}^{-1}_{c(i),j}$. The dictionaries $\{\mathbf{D}_{c(i),j}: i=1,2,\cdots,N_p, \, j=1,2\}$ are parameterized by indexes $i$ and $j$, meaning that the dictionaries are patch dependent and also that the numerical gradients along the horizontal and vertical directions are handled separately. As we will see later, patches are grouped into several major classes, and a single dictionary will take care of one group. Therefore we use a many-to-one function $c(i)$ to describe this relationship: $c(i) = s, \, \forall i \in I_s$, where $I_s$ is the index set of patches belonging to group $s$. 

The dictionaries can be constructed before the inversion with an external database of patterns that we think are close to the geological problems at hand, or they can be learned during the inversion process in a bootstrapping way, where only internal information is used. In this study, we take the second approach; that is, we train the dictionaries using the inversion result from the previous iteration, and use those dictionaries to guide the inversion in the current iteration. We note that a dictionary $\mathbf{D}_{c(i),j}$ is also a function of model $\mathbf{m}$, which means that the dictionaries change with iterations, but we do not explicitly write this out to avoid cluttering of symbols. Finally, the $\ell_1$ norm forces the coefficients to be sparse. 

In summary, the regularization requires that every patch extracted from the model derivatives be sparse under the representation of some corresponding dictionary. One can see the direct connection between the proposed regularization and total-variation regularization (See Appendix \ref{sec:TV}). In fact, if the patch extraction operator extracts the whole model, and the dictionaries are identity matrices, then the NMAS regularization reduces to total variation regularization.


\noindent \textbf{Algorithm}

Applying the Alternating Direction Method of Multipliers (ADMM) described in the Appendix \ref{sec:admm}, we first convert the original problem into an equality constrained problem
\begin{equation}
\label{eq:objective_equal_constraint}
\begin{split}
& \argmin_{\mathbf{m}, \bm{\alpha}} \frac{1}{2}\|\mathbf{d} - f(\mathbf{m})\|^2_2 
+ \beta \sum_{i,j}\|\bm{\alpha}_{i,j}\|_1, \\
& \text{ s.t. } \mathbf{R}_{i}\bm{\nabla}_j \mathbf{m} = \mathbf{D}_{c(i),j}\bm{\alpha}_{i, j}, \, j = 1,2, \forall i
\end{split}
\end{equation}

The augmented Lagrangian is formulated as
\begin{equation}
\label{eq:AL}
\begin{split}
	\mathcal{L_{\mathcal{A}}(\mathbf{m}, \bm{\alpha}}) = 
	& \frac{1}{2} \|\mathbf{d} - f(\mathbf{m})\|_2^2 + \beta\sum_{i,j}\|\bm{\alpha}_{i,j}\|_1 + \sum_{i,j}\bm{\gamma}_{i, j}^T(\mathbf{R}_{i}\bm{\nabla}_j \mathbf{m} - \mathbf{D}_{c(i),j}\bm{\alpha}_{i,j}) \\
	& + \frac{\rho}{2}\sum_{i,j}\|\mathbf{R}_{i}\bm{\nabla}_j \mathbf{m} - \mathbf{D}_{c(i),j}\bm{\alpha}_{i,j}\|_2^2,
\end{split}
\end{equation}
where the vector $\bm{\alpha}$ on the left-hand side is the concatenation of all sparse coefficient vectors $\bm{\alpha}_{i,j}$ on the right-hand side, $\bm{\gamma}_{i, j}$ are the Lagrange multipliers, the last term is a quadratic penalty term, and $\rho$ is the penalty parameter, which is the step size in the dual update. We will discuss how to choose $\rho$ and $\beta$ in practice in Section \ref{sec:algorithm}. Compared to a standard Lagrangian, the augmented Lagrangian has an extra quadratic penalty term. Compared to penalty methods, the augmented Lagrangian has an extra term of Lagrange multipliers. The theoretical details of the benefits of using an augmented Lagrangian with respect to the other two methods are beyond the scope of this paper. Interested readers are referred to \cite{boyd2011distributed} and \cite{nocedal2006numerical}.

To make the optimization easier, we rewrite Problem (\ref{eq:AL}) in the scaled form \citep{boyd2011distributed},
\begin{equation}
\label{eq:SAL}
\begin{split}
	\mathcal{L_{\mathcal{A}}(\mathbf{m}, \bm{\alpha}, \mathbf{v}}) = 
	& \frac{1}{2} \|\mathbf{d} - f(\mathbf{m})\|_2^2 + \beta\sum_{i,j}\|\bm{\alpha}_{i,j}\|_1  \\
	& + \frac{\rho}{2}\sum_{i,j}\|\mathbf{R}_{i}\bm{\nabla}_j \mathbf{m} - \mathbf{D}_{c(i),j}\bm{\alpha}_{i,j} + \mathbf{v}_{i,j}\|_2^2 + \text{const},
\end{split}
\end{equation}
where $\mathbf{v}_{i,j} = (1/\rho)\bm{\gamma}_{i,j}$, which are the scaled Lagrange multipliers.

The ADMM iteration scheme derived from the scaled augmented Lagrangian is:
\begin{equation}
\label{eq:m_update}
	\mathbf{m}^{k+1} := \argmin_{\mathbf{m}} \frac{1}{2} \|\mathbf{d} - f(\mathbf{m})\|_2^2 + \frac{\rho}{2}\sum_{i,j}\|\mathbf{R}_{i}\bm{\nabla}_j \mathbf{m} - \mathbf{D}^k_{c(i),j}\bm{\alpha}_{i,j}^k + \mathbf{v}_{i,j}^k\|_2^2,
\end{equation}
\begin{equation}
\label{eq:alpha_update}
	\bm{\alpha}^{k+1} := \argmin_{\bm{\alpha}} \frac{\rho}{2}\sum_{i, j} \|\mathbf{R}_{i}\bm{\nabla}_j\mathbf{m}^{k+1} - \mathbf{D}^{k+1}_{c(i),j}\bm{\alpha}_{i, j} + \mathbf{v}_{i,j}^k\|^2_2 + \beta\|\bm{\alpha}_{i,j}\|_1,
\end{equation}
\begin{equation}
\label{eq:v_update}
	\mathbf{v}_{i,j}^{k+1} = \mathbf{v}_{i,j}^{k} + (\mathbf{R}_{i}\bm{\nabla}_j \mathbf{m}^{k+1} - \mathbf{D}^{k+1}_{c(i),j}\bm{\alpha}_{i,j}^{k+1}).
\end{equation}
We loop through the above three sub-problems until convergence or reaching the maximum allowed iteration numbers. In each sub-problem, we keep other parameters fixed and find the update of the argument of the current problem.

For Problem (\ref{eq:m_update}), the gradient of the objective function with respect to $\mathbf{m}$ is
\begin{equation}
\label{eq:gradient_m_problem}
	\nabla_{\mathbf{m}}\chi_1 = -\left(\frac{\partial f}{\partial \mathbf{m}}\right)^T\left(\mathbf{d} - f(\mathbf{m})\right) + \rho \sum_{i,j}\bm{\nabla}_j^T (\mathbf{R}^T_i \mathbf{R}_i\bm{\nabla}_j\mathbf{m} - \mathbf{R}^T_i \mathbf{D}^{k}_{c(i),j}\bm{\alpha}_{i,j}^k + \mathbf{R}^T_i \mathbf{v}_j^k),
\end{equation}
so it can be solved using the L-BFGS algorithm mentioned before. Note that $\sum_{i}\bm{\nabla}_j^T \mathbf{R}^T_i \mathbf{R}_i\bm{\nabla}_j\mathbf{m} = \bm{\nabla}_j^T \left(\sum_{i} \mathbf{R}^T_i \mathbf{R}_i \right)\bm{\nabla}_j\mathbf{m} = n^2\bm{\nabla}_j^T \bm{\nabla}_j\mathbf{m}$, because of our previous selection of stride and boundary conditions and the linearity of $\bm{\nabla}_j$, which make the computation easier.

The Problem (\ref{eq:alpha_update}) is equivalent to
\label{eq: alpha_update2}
\begin{equation}
	\bm{\alpha}_{i,j}^{k+1} := \argmin_{\bm{\alpha}_{i,j}} (1/2)\left\Vert \left( \mathbf{R}_i\bm{\nabla}_j \mathbf{m}^{k+1} + \mathbf{v}_j^k\right) - \mathbf{D}^{k+1}_{c(i),j}\bm{\alpha}_{i,j}\right\Vert_2^2 + (\beta/\rho)\|\bm{\alpha}_{i,j}\|_1,
\end{equation}
for every $i$ and $j$ since the problem is separable. This problem has a closed-form solution by soft-thresholding:
\begin{equation}
\label{eq: alpha_update3}
	\bm{\alpha}_{i,j}^{k+1} := S_{\beta/\rho} \left( \left({\mathbf{D}^{k+1}_{c(i),j}}\right)^T\left( \mathbf{R}_i\bm{\nabla}_j \mathbf{m}^{k+1} + \mathbf{v}_j^k\right) \right).
\end{equation}

Between Problem (\ref{eq:m_update}) and Problem (\ref{eq:alpha_update}), we update the dictionaries by solving
\begin{equation}
\label{eq:mclass-dictionary}
	\argmin_{\mathbf{D}_{s,j}, \mathbf{X}} \|\mathbf{Y}_{s,j}^{k+1} - \mathbf{D}_{s,j} \mathbf{X}\|_F^2 + \lambda \sum_l \|\mathbf{X}(:,l)\|_1, \, \text{s.t. } \mathbf{D}_{s,j}^T \mathbf{D}_{s,j} = \mathbf{I},
\end{equation}
where $\mathbf{Y}_{s,j}^{k+1}$ stands for the matrix of training patch vectors obtained from patch subset $s$ for derivative direction $j$. We will talk about the multi-class dictionary learning procedures and the significance of the steps in the iteration scheme in section \ref{sec:mcd}.

\noindent \textbf{Interpretation}

To further understand how this iterative scheme works, we interpret it roughly as FWI plus an artifact removal process with the nonlocal similarity prior implemented by multi-class dictionaries. In fact, Problem (\ref{eq:m_update}) is a FWI problem with a quadratic penalty term. The penalty term requires that the model derivative patches be close to certain variables. Those variables are actually the sparse approximations of the model derivative patches from the previous iteration, plus dual variables derived from the algorithm. As mentioned in section \ref{sec:dl}, solving the sparse approximation problem (\ref{eq:alpha_update}) by soft-thresholding reduces artifacts and in the meantime preserves small-scale geological structures.

\begin{figure}
\begin{center}
\includegraphics[width=0.8\textwidth]{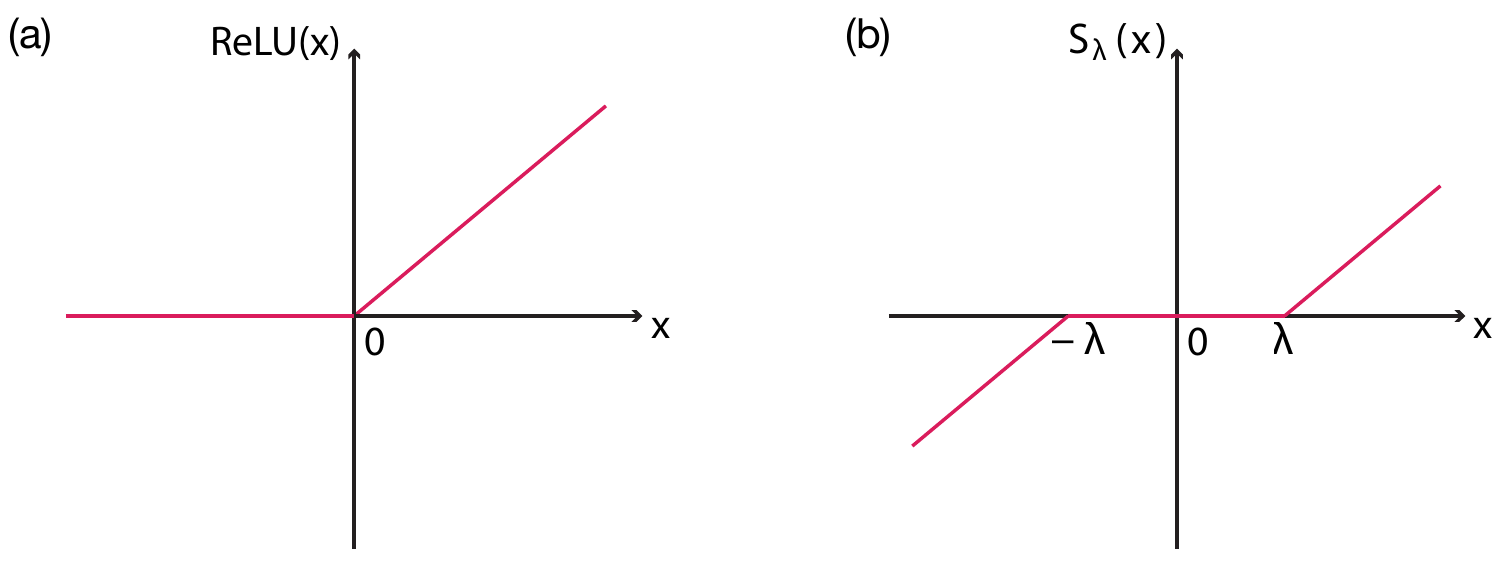}
\end{center}\vspace{-0.5cm}
\caption{Illustration of ReLU and the soft-thresholding operator. (a). ReLU ; (b). Soft-thresholding operator $S_{\lambda}(x)$. }
\label{fig:relu_soft}
\end{figure}

Inspired by \cite{dong2014learning}, we find a convolutional neural network (CNN) \citep{lecun1989backpropagation,goodfellow2016deep} hidden in the above artifact removal process, whose layers are constructed individually with prior knowledge. In fact, through sparse approximations the inverted model from the $k$-th iteration contributes to the FWI gradient in Equation (\ref{eq:gradient_m_problem}) as 
\begin{equation}
\mathbf{g} = -\sum_{i}\bm{\nabla}^T \mathbf{R}^T_i \mathbf{D}_{c(i)} S_{\beta/\rho} ( \mathbf{D}^T_{c(i)} \mathbf{R}_i\bm{\nabla} \mathbf{m}^{k} ).
\end{equation} 
Here we ignore the dual variable $\mathbf{v}_j^k$, drop the subscript $j$ and only consider the vertical direction for the convenience of interpretations. The finite-difference filter in the first layer detects low-level features such as edges. Then, the patch extraction operator and transposed dictionaries act as a convolutional layer with periodic padding, where each filter is an atom from the dictionaries. Different from those in conventional CNNs, these filters can be regarded as spatially variable, as patches at different locations require different dictionaries. Following that, the output goes through soft-thresholding -- a nonlinear mapping that is similar to the ReLU layer (Figure \ref{fig:relu_soft}) commonly used in CNNs \citep{glorot2011deep}. The remaining layers can be interpreted similarly as transposed convolutional layers. In short, the regularization is unrolled as the following process: an intermediate inverted model is fed into a special CNN, and the output is added to the FWI gradient to nudge the inversion towards a direction that would satisfy our prior knowledge.

\subsection{Nonlocal Similarity}
\label{subsec:NLS}
Within the same model, one patch may recur multiple times at different locations, both within the same scale or across different scales, under rotations and shearing. In other words, we consider patches to be similar if equivalent under an affine transformation. This simple yet powerful prior of nonlocal similarity was first introduced by the image processing community, and has been widely used in image denoising \citep{buades2005review,dabov2007image}, image restoration and super-resolution tasks \citep{dong2011image,glasner2009super,huang2015single}. 

How do we quantify the similarity? Although the commonly-used patch-wise Euclidean distance or cross-correlation can be used as a measure for this purpose, we find that they tend to produce poor results. A better choice is the distance between image feature descriptors. Those feature descriptors are compact vectors encoding characteristic feature information that we are interested in the patch. In our algorithm, we adopt the histogram of oriented gradients (HoG) descriptor \citep{dalal2005histograms}. As its name suggests, the HoG feature descriptor constructs a histogram of the gradient orientations of an image patch. The value in each bin is the sum of the magnitudes of gradients falling in the orientation range of that bin.

In our implementation, we first take two derivative filters $f_v = \left[-1, 0, 1\right]^T$ and $f_h = \left[-1, 0, 1\right]$, and convolve them with the patch to get gradients at each point along the vertical and horizontal directions, denoted by $\mathbf{g}_v(x,y)$ and $\mathbf{g}_h(x,y)$, respectively. Then we compute the gradient magnitude as $\lvert \mathbf{g}(x,y) \rvert = \sqrt{\mathbf{g}_v^2(x,y) + \mathbf{g}_h^2(x,y)}$, and the angle as $\angle \mathbf{g}(x,y) = \tan^{-1} \left[\mathbf{g}_h^2\left(x,y\right) / (\mathbf{g}_v^2\left(x,y\right) + \epsilon)\right]$, where $\epsilon$ is a small number to prevent division by zero. The angle ranges from $-90^{\circ}$ to $90^{\circ}$, and is divided into 9 bins (here we follow the settings from \citep{dalal2005histograms}), which are $(-90^{\circ}, -70^{\circ}, -50^{\circ}, -30^{\circ}, -10^{\circ}, 10^{\circ}, 30^{\circ}, 50^{\circ}, 70^{\circ})$. The $0^{\circ}$ direction points vertically downward. For each point in the patch, we find the two closest bins and split its gradient magnitude into the two bins, proportional to its inverse distance to the two bins. Repeating the procedure for all points and summing their votes, we obtain a description vector. We then normalize it to produce our HoG feature descriptor of this patch. Note that the patches here are actually the derivatives of velocity patches mentioned in previous sections.



\begin{figure}
\begin{center}
\includegraphics[width=15cm]{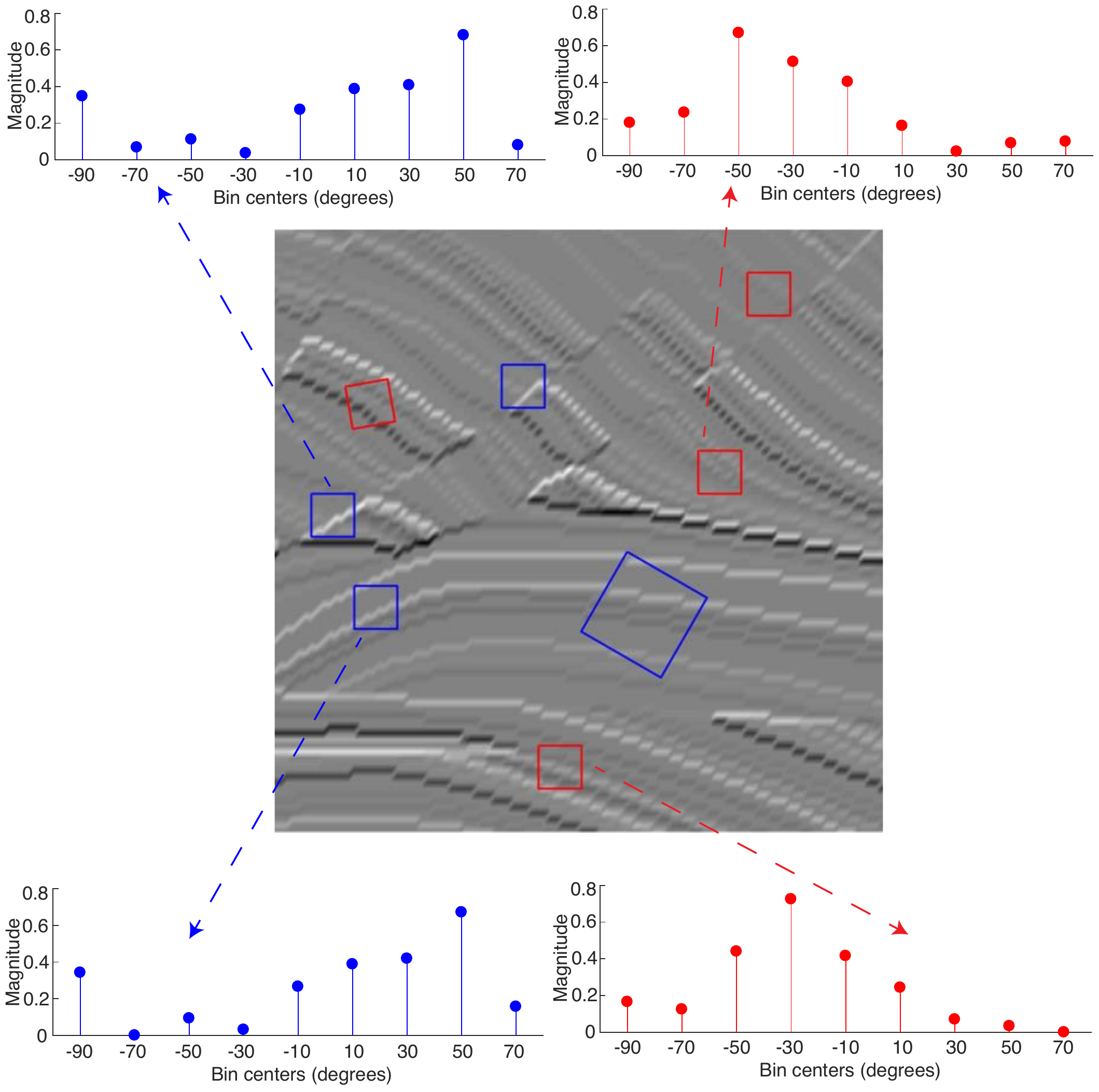}
\end{center}\vspace{-0.5cm}
\caption{Illustration of nonlocal similarity. We show the vertical derivative of a small section cropped from the Marmousi model. The color is close to gray if the values are close to 0, so the derivative field is sparse. We show patch examples belonging to two different groups, bounded by blue and red boxes respectively. Structures in the blue boxes dip to the left, while those in the red boxes dip to the right. Within each group, structures of different scales and orientations are similar to each other, as is demonstrated by their HoG descriptors.}
\label{fig:nl_similarity}
\end{figure}

We measure the nonlocal similarity in the model-derivative domain. Figure \ref{fig:nl_similarity} shows the vertical derivative of the central part of the Marmousi model. Two groups of patches are respectively bounded by blue and red boxes of different sizes and orientations. The ones in the blue boxes dip to the left, while those in the red boxes dip to the right. Within each group, patches are similar to each other, demonstrated by the similarity of their corresponding HoG feature descriptors.

In the research works mentioned at the beginning of this subsection, the nonlocal similarity regularization term usually penalizes the difference between each patch and the linear combinations of patches similar to it, with those coefficients for linear combinations proportional to the patch-wise similarity. For a given gradient direction $j$, we have
\begin{equation}
 	\mathcal{C}_{\mathbf{m}_1}(\mathbf{m}) = \sum_i \|\mathbf{R}_{i}\bm{\nabla}_j \mathbf{m} - \mathbf{T}_{i,j}(\bm{\nabla}_j \mathbf{m}) \mathbf{w}_{i,j}\|_2^2,
\end{equation}
where $\mathbf{T}_{i,j}: \mathbb{R}^{MN} \rightarrow \mathbb{R}^{n^2\times p}$ is an operator that extracts the $p$ most similar patch to the target patch with index $i$ and pack them as column vectors in a matrix, and $\mathbf{w}_{i,j}$ is the corresponding weight vector. Both $\mathbf{T}_{i,j}$ and $\mathbf{w}_{i,j}$ depend on $\mathbf{m}$, the unknown model vector that we try to solve for, so the regularization term is nonlinear.

Now we show that the iterative scheme in section \ref{sec:objfunc} corresponds to nonlocal similarity regularization.

We assume that patches are similar up to affine transformations, so the search space is very large and it is an intimidating task to find the $p$ most similar patches for every single patch. Therefore, we use a strategy similar to \cite{mairal2009non}, where we first extract patches of different scales and orientations across the model, and then group similar ones into disjoint sets. For each set of patches, we use dictionary learning to construct a dictionary whose atoms can compactly represent the structure patterns of this category. After this dictionary learning step, the regularization term becomes
\begin{equation}
 	\mathcal{C}_{\mathbf{m}_2}(\mathbf{m}) = \sum_i \|\mathbf{R}_{i}\bm{\nabla}_j \mathbf{m} - \mathbf{D}_{c(i),j}\bm{\alpha}_{i,j}\|_2^2,
\end{equation}
where the dictionaries $\mathbf{D}_{c(i),j}$ are like $\mathbf{T}_{i,j}(\bm{\nabla}_j \mathbf{m})$, and the sparse coefficients $\bm{\alpha}_{i,j}$ act as $\mathbf{w}_{i,j}$. Note that here $\mathbf{D}_{c(i),j}$ and $\mathbf{w}_{i,j}$ also depend on the unknown model vector $\mathbf{m}$. To handle the nonlinearity we decompose the regularization into three steps. Suppose we already obtain the model at iteration $k$. First, we construct dictionaries $\mathbf{D}_{s,j}^k$ for each subset of derivative patches from $\mathbf{m}^k$:
\begin{equation}
\label{eq:mclass-dictionary2}
	\argmin_{\mathbf{D}_{s,j}, \mathbf{X}} \|\mathbf{Y}_{s,j}^k - \mathbf{D}_{s,j} \mathbf{X}\|_F^2 + \lambda \sum_l\|\mathbf{X}(:,l)\|_1, \, \text{s.t. } \mathbf{D}_{s,j}^T \mathbf{D}_{s,j} = \mathbf{I},
\end{equation}
just as Problem (\ref{eq:mclass-dictionary}). Then, we compute $\bm{\alpha}_{i,j}$ by solving a $\ell_1$-minimization problem, also with $\mathbf{m}^k$:
\begin{equation}
 	\bm{\alpha}^{k} := \argmin_{\bm{\alpha}} \sum_{i, j} \|\mathbf{R}_{i}\bm{\nabla}_j\mathbf{m}^{k} - \mathbf{D}^k_{c(i),j}\bm{\alpha}_{i, j}\|^2_2 + \tilde{\beta}\|\bm{\alpha}_{i,j}\|_1,
\end{equation}
where $\bm{\alpha}^k$ is the concatenation of all coefficient vectors $\bm{\alpha}_{i,j}^k$. This is the same as the ``$\bm{\alpha}$-problem'' (Problem \ref{eq:alpha_update}), except for the scaled Lagrange multipliers $\mathbf{v}_{i,j}^k$. With the dictionaries and sparse coefficients, we let the regularization term for $\mathbf{m}_{k+1}$ be
\begin{equation}
 	\mathcal{C}_{\mathbf{m}_2}(\mathbf{m}) = \sum_{i,j}\|\mathbf{R}_{i}\bm{\nabla}_j \mathbf{m} - \mathbf{D}^k_{c(i),j}\bm{\alpha}_{i,j}^k\|_2^2.
\end{equation}
We see that this is the same as the regularization term in the ``$\mathbf{m}$-problem'' (Problem (\ref{eq:m_update})). At this point, it is clear that the proposed regularization defined in objective function (\ref{eq:objective}) is essentially a nonlocal similarity regularization on top of total variation.

\subsection{Multi-class Orthogonal Dictionary Learning}
\label{sec:mcd}
In this subsection, we describe how to solve the multi-class orthogonal dictionary learning problem (\ref{eq:mclass-dictionary} or \ref{eq:mclass-dictionary2}).

\begin{figure}
\begin{center}
\includegraphics[width=16cm]{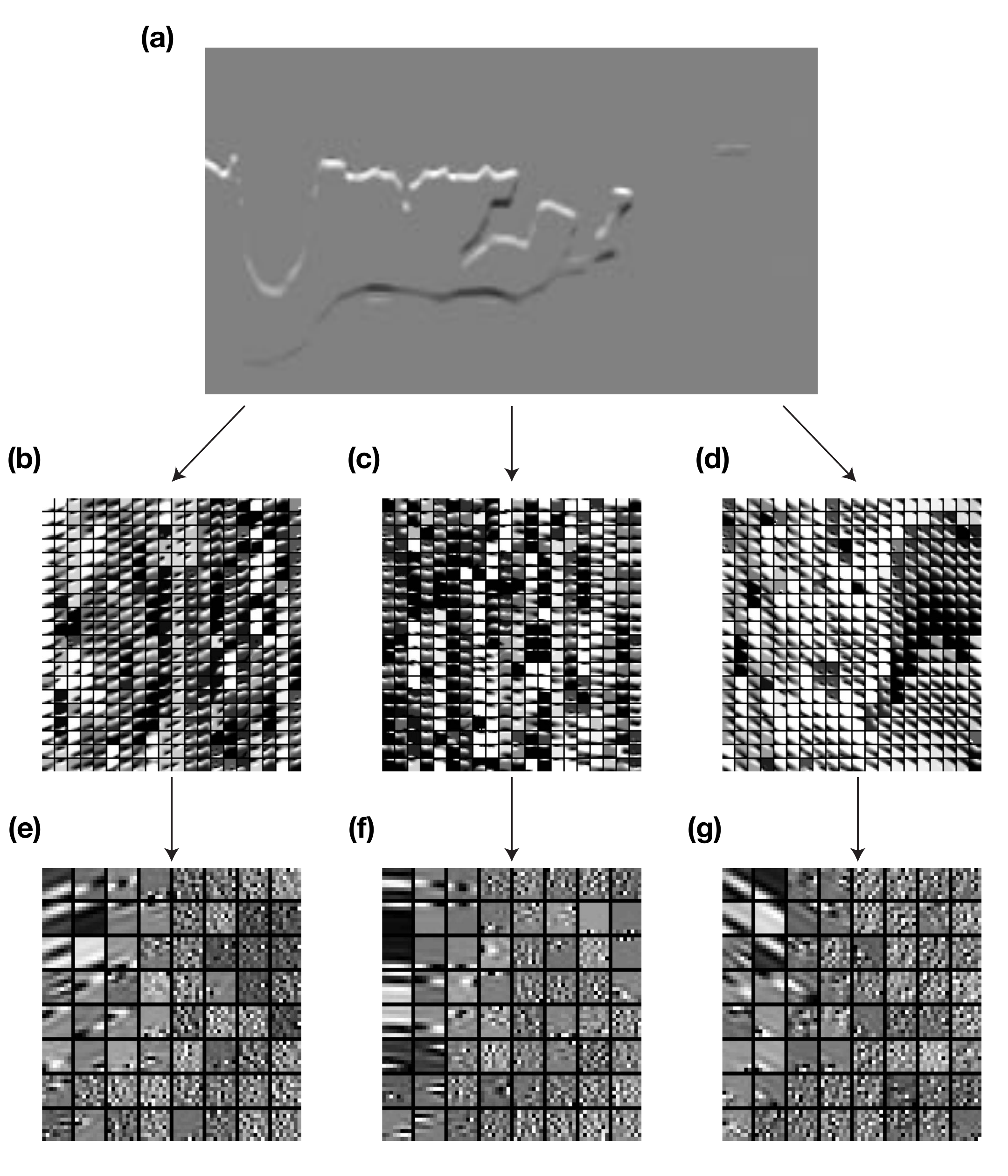} \\[-0.7cm]
\end{center}\vspace{-0.15cm}
\caption{Three different classes of training patches and corresponding dictionaries. (a) The vertical derivative for patch extraction. (b), (c), and (d) are first 400 patches from three typical clusters. Patches in (b) dip to the left, patches in (c) are almost blocks and are horizontal, and patches in (d) dip the the right. (e), (f), and (g) are orthogonal dictionaries trained from corresponding patch clusters. One can clearly see that the atoms in those dictionaries capture the dipping information.}
\label{fig:trainPatchClass}
\end{figure}

For each class of patches, the orthogonal dictionary learning procedure is the straightforward iterative scheme described by Problem (\ref{eq:dicttion_learning_X}) and (\ref{eq:dicttion_learning_D}). Therefore, the multi-class dictionary learning problem actually boils down to how to construct $\mathbf{Y}_{s,j}^k$, which are the matrices of training patch vectors, where $s$ stands for the indices of patch sets and $j$ stands for model derivative directions.

With the notations from the previous section, we assume that we have the model vector at the $k$th iteration after solving Problem (\ref{eq:m_update}). Problem (\ref{eq:alpha_update}) suggests that $(R_{i}\bm{\nabla}_j \mathbf{m}^k+\mathbf{v}_{i,j}^{k-1})$ are the target collection of patches for which we need to find a sparse approximation (the scaled Lagrange multipliers are absorbed into the model patches). Moreover, following Appendix \ref{sec:admm} we see that in the ADMM iteration scheme for the TV regularized FWI, the gradient field is soft-thresholded to be sparse, that is, the solution to
\begin{equation}
	\mathbf{z}_i^{k+1} := \argmin_{\mathbf{z}_i} \beta\|\mathbf{z}_i\|_1 + (\rho/2)\|\bm{\nabla}_i \mathbf{m}^{k+1} - \mathbf{z}_i + \mathbf{u}^k_i\|^2_2
	\label{eq:TV_z_update}
\end{equation}
is
\begin{equation}
	\mathbf{z}_i^{k+1} = S_{\beta/\rho} \left(\bm{\nabla}_i \mathbf{m}^{k+1} + \mathbf{u}^k_i \right).
\end{equation}

Based on the above analysis, we can use 
\begin{equation}
\left(\sum_i \mathbf{R}_i^T \mathbf{R}_i\right)^{-1}\sum_{i} \mathbf{R}_i^T S_{\beta/\alpha}(\mathbf{R}_i \bm{\nabla}_j \mathbf{m}^k + \mathbf{v}_{i,j}^{k-1}) = (1/n^2)\sum_{i} \mathbf{R}_i^T S_{\beta/\alpha}(\mathbf{R}_i \bm{\nabla}_j \mathbf{m}^k + \mathbf{v}_{i,j}^{k-1}),
\end{equation}
to extract patches from, so that we can combine TV and our method and make the training patches sparser and have less artifacts.

Next we extract every possible patch for a given range of scales and orientations to form a large patch collection. To be more specific, by different scales we mean that before patch extractions we shrink the model by 1, 0.75, and 0.5, for example. In this way, a fixed-size sliding window can crop patterns of larger scales in a shrinked model, incorporating multi-resolution information. Besides, the sliding window can be rotated by different angles.


After that, we need to divide the collection of patches into subsets to train those dictionaries. This requires us to group similar patches together into separate clusters using a certain criterion. In our algorithm, we opt to group the training patches according to the geometrical directions of their patterns, which is a similar strategy to that of a plane-wave destruction filter \citep{claerbout1992earth,fomel2002applications} or geologically constrained migration velocity analysis \citep{Bob_clapp_thesis}. Thus, we adopt the HoG feature descriptor described in subsection \ref{subsec:NLS}, use the $\ell_2$ norm of their difference as the measure of similarity between patches, and adopt the k-means++ method \citep{arthur2007k} to group those patches into $k$ clusters.


Then, we use the orthogonal dictionary learning introduced in the previous section to construct a dictionary for each cluster. Figure~\ref{fig:trainPatchClass} shows examples of three classes of training patches and their corresponding dictionaries.

As a side note, the multi-class orthogonal dictionaries have another advantage. Usually, one orthogonal dictionary alone is not as expressive as an overcomplete dictionary like K-SVD due to its relatively small number of atoms; thus, it may not yield as sparse an approximation as as the latter. This can be understood from another angle. The whole training set of patches may include so many patterns that a single orthogonal dictionary cannot encode them all very well. However, this ensemble of multi-class dictionaries can be very expressive, since one single dictionary from it can handle one class of patterns. Also, we still enjoy the advantage of orthogonal dictionaries such as high learning speed, and the ability to use simple soft-thresholding rather than more expensive orthogonal matching pursuit (OMP) \citep{pati1993orthogonal} in Problem (\ref{eq:alpha_update}).



\subsection{Algorithm Summary}
\label{sec:algorithm}
\begin{figure}
\begin{center}
\includegraphics[width=16cm]{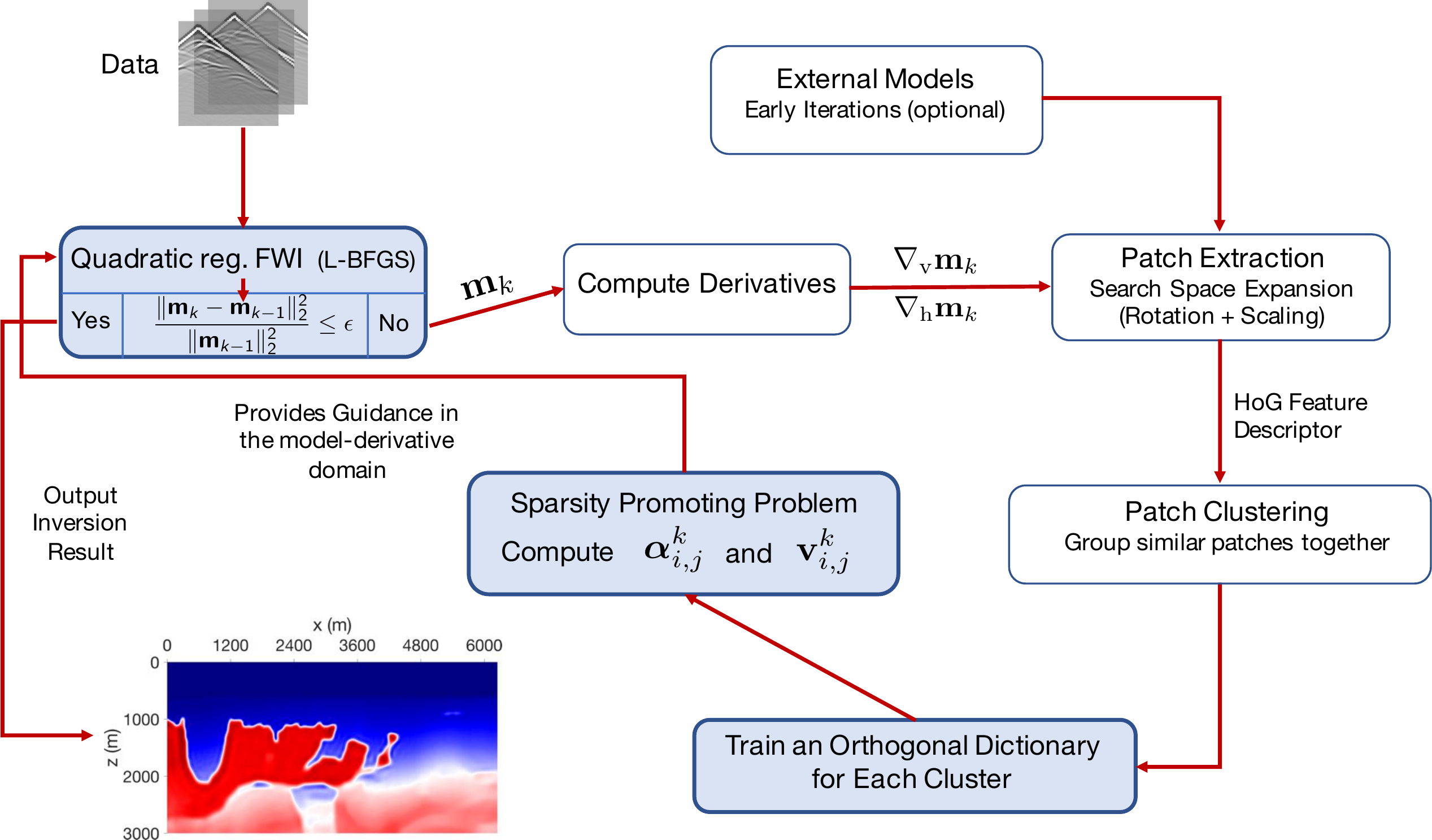} \\[-0.7cm]
\end{center}\vspace{-0.15cm}
\caption{Flowchart of the NMAS algorithm.}
\label{fig:flowchart}
\end{figure}

The algorithm of the NMAS method developed to solve Problems (\ref{eq:m_update}), (\ref{eq:mclass-dictionary}), (\ref{eq:alpha_update}), and (\ref{eq:v_update}) is summarized in the flowchart shown in Figure \ref{fig:flowchart}.

Following this flowchart, we now go through the algorithm in detail.

\begin{enumerate}
\item[(1).] Before inversion, we initialize the weighting factors $\rho$ and $\beta$, set the coefficients $\bm{\alpha}_{i,j}^0 = \mathbf{0}$, and the scaled Lagrange multipliers $\mathbf{v}_{i,j}^0 = \mathbf{0}$. Also, $\mathbf{m}_0$ is set to be the initial model.

\item[(2).] The algorithm consists of two levels of loops. Embedded in the outer loop shown in the flowchart, the inner loop is shown in the first block after input of data -- the FWI with a quadratic regularization term which is Problem (\ref{eq:m_update}). We solve this inner loop problem by the L-BFGS algorithm \citep{nocedal2006numerical}. The algorithm stops if the maximum number of iterations is reached, or the objective function decrease is small enough ($\leq$ 1e-9), or the infinity norm of the gradient is small enough ($\leq$ 1e-5). After each pass of the inner loop, we obtain a new model $\mathbf{m}_{k}$, and continue iterating in the outer loop if the relative error between two iterations $\|\mathbf{m}_{k} - \mathbf{m}_{k-1}\|^2_2/\|\mathbf{m}_{k-1}\|^2_2$ is still larger than the given threshold $\epsilon$. In practice, we can also terminate the inversion when the number of iterations of the outer loop reaches a predefined maximum value. In the first iteration of the outer loop, we use the inner loop FWI with no regularization. At the second iteration of the outer loop, before entering the inner loop block, we determine $\rho$ and $\beta$ by empirical formulas as in \citet{guitton2012blocky}:
\begin{equation}
	R_\rho = \left( \frac{\rho}{2}\sum_{i,j}\|\mathbf{R}_{i}\bm{\nabla}_j \mathbf{m} - \mathbf{D}_{c(i),j}\bm{\alpha}_{i,j} + \mathbf{v}_{i,j}\|_2^2 \right) / \left( \frac{1}{2}\|\mathbf{d} - f(\mathbf{m})\|^2_2 \right),
\end{equation}
and 
\begin{equation}
	R_\beta = \left( \beta\sum_{i,j}\|\bm{\alpha}_{i,j}\|_1 \right) / \left( \frac{1}{2}\|\mathbf{d} - f(\mathbf{m})\|^2_2 \right),
\end{equation}
where $R_\rho$ and $R_\beta$ are preset parameters, for which values of the order of $10^{-3}$ work well.

\item[(3).] Once we find a new model, we proceed to compute their vertical and horizontal derivatives and feed them into patch extraction operations, where we expand the search space by cropping patches of difference sizes and orientations. At this step, we can do extra processing on the patches. For example, we may select patches with high variance -- those with strong geological patterns. We can also threshold the patches like the sparse-approximation step in TV regularization to further reduce artifacts and noises. Since we use the nonlocal similarity prior, the patches are extracted from internal models. However, it is also possible to introduce information outside of the inversion by adding in external models. In fact, outside information may guide the inversion in a direction of faster convergence, especially in early iterations.

\item[(4).] With these vast number of patches, we use HoG feature descriptor to create a compact representation for each of them, and then use clustering algorithms to group similar patches together.

\item[(5).] We find an orthogonal dictionary for each cluster following Problem (\ref{eq:mclass-dictionary}), and then solve the sparsity promoting problem (\ref{eq:alpha_update}). Also, we update the scaled Lagrange multipliers $\mathbf{v}_{i,j}^k$ by Problem (\ref{eq:v_update}). These steps generate $\bm{\alpha}_{i,j}^k$ and $\mathbf{v}_{i,j}^k$, which are again substituted into the inner loop, to provide guidance for the next round of FWI. Once the termination condition is satisfied, the algorithm outputs the final result.

\end{enumerate}

We also mention two strategies adopted for the ADMM iteration scheme here \citep{boyd2011distributed}. The first one is a ``warm start", meaning that for the FWI in the inner loop, we use the model obtained from the last iteration in the outer loop as the starting model, to speed up convergence. The second strategy is called ``Inexact minimization'': we gradually increase the maximum allowed number of iterations in the inner loop as the number of outer loop iteration increases. This is based on the consideration that early inner iterations are not accurate so it is sufficient to solve them approximately at first, and gradually increase the accuracy along the way. We can save computational cost with this strategy.

\section{Inversion Results}
\subsection{Modified BP 2004 Model}
\begin{figure}
\begin{center}
\includegraphics[width=\textwidth]{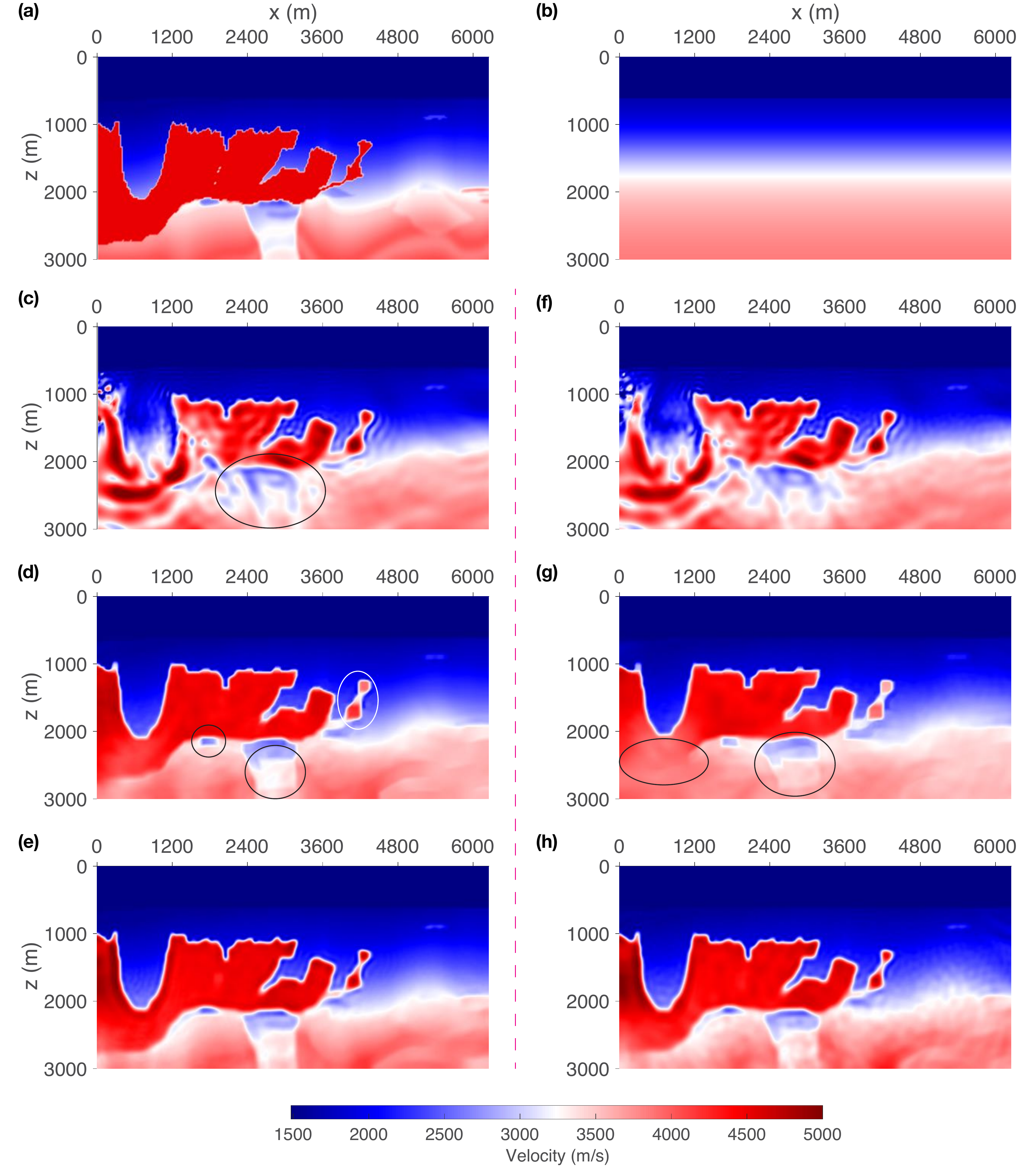}
\end{center}\vspace{-0.5cm}
\caption{The FWI inversion results of the BP 2004 model. (a). The true velocity model; (b). The 1-D initial model; (c). The inversion result without regularization; (d) The inversion result with TV regularization; (e). The inversion result with NMAS regularization; (f). The inversion result without regularization with added noise in the data; (g). The inversion result with TV regularization with added noise; (h). The inversion result with NMAS with added noise. Comparing (c), (d), (e) or (f), (g), and (h), one can see that the salt body is best delineated and the sub-salt velocity is the most accurately resolved using NMAS.}
\label{fig:BP_results}
\end{figure}

We use a modified version of the BP 2004 velocity model \citep{billette20052004} for numerical tests. The model is first cropped from the original one from 0m to 32150m in the horizontal direction and from 0m to 7500m in depth. The model is then decimated in grid points to save computation time and scaled to the dimension of 3000m $\times$ 6250m, with a grid spacing of 25m. The sea-bottom bathymetry would be inaccurate due to the subsampling, so we replace the top 600m part of the scaled model with a homogeneous water layer. The depth of the salt dome is kept the same in this way. The model is shown in Figure~\ref{fig:BP_results}(a). The initial model used to start the FWI is 1-D, obtained by first smoothing the slowness of the modified model, and then averaging along the horizontal direction (Figure~\ref{fig:BP_results}(b)). The major difficulties of inversion for this model are salt delineation and accurate reconstruction of sub-salt structures.

\begin{figure}
\begin{center}
\includegraphics[width=0.8\textwidth]{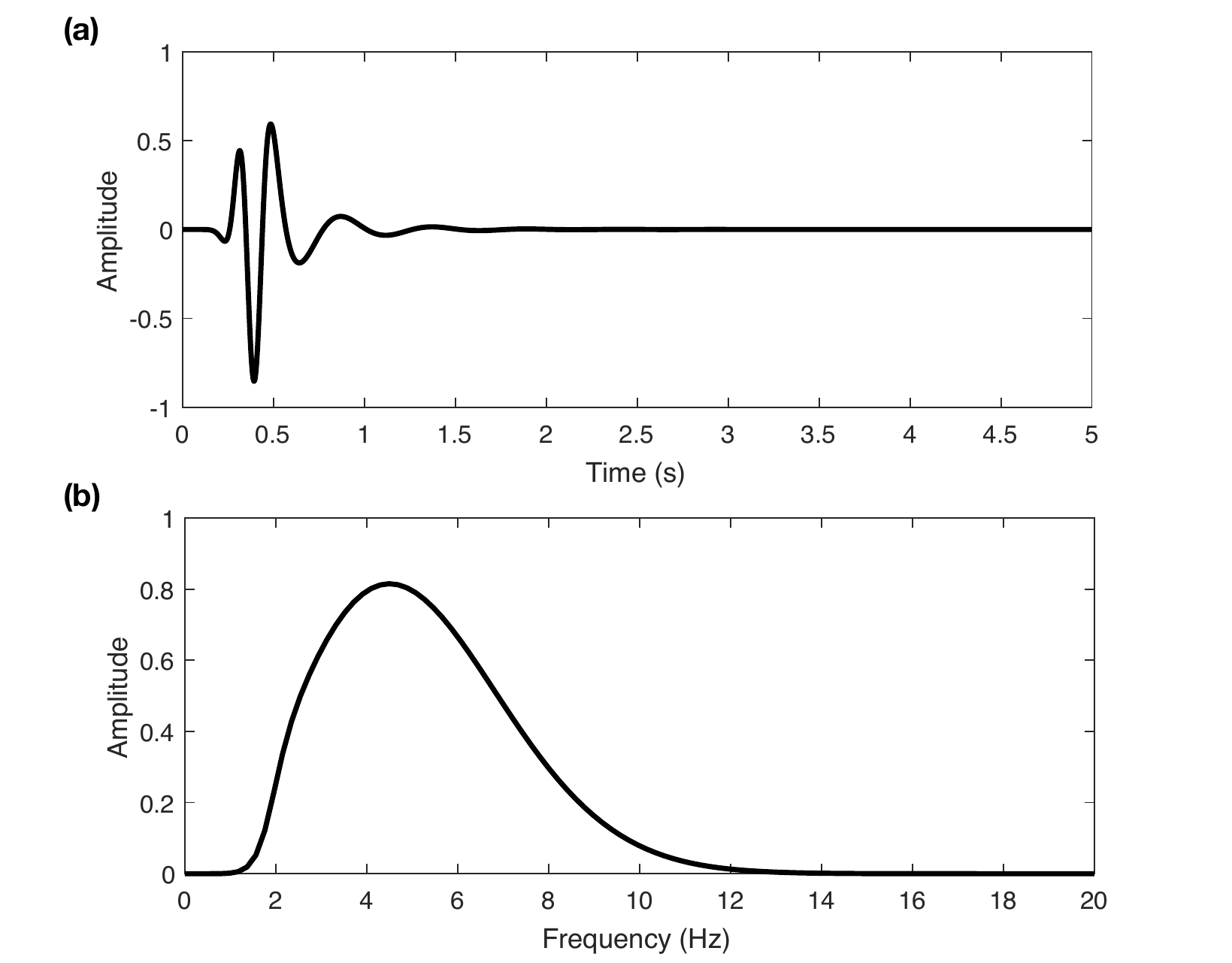}
\end{center}\vspace{-0.5cm}
\caption{Source-time function with a dominant frequency of 4.5Hz and its spectrum. (a). The source-time function; (b). The spectrum. The frequency contents below 2Hz are attenuated.}
\label{fig:Source_time_function}
\end{figure}

\begin{figure}
\begin{center}
\includegraphics[width=0.8\textwidth]{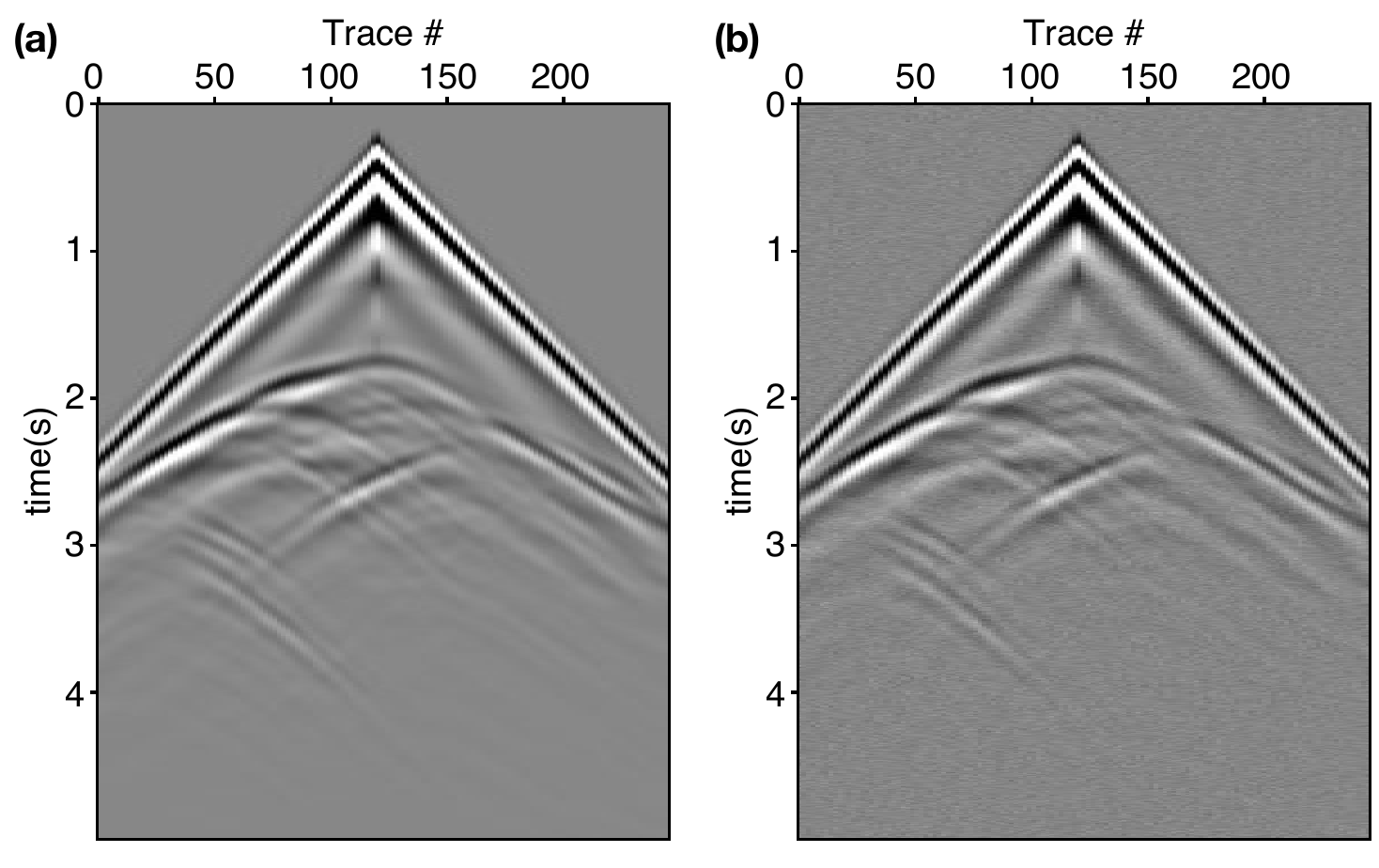}
\end{center}\vspace{-0.5cm}
\caption{Examples of the data. (a). Shot gather No. 25 without noise; (b). Shot gather No. 25 with noise.}
\label{fig:Filtered_data}
\end{figure}

In this experiment, we use a fix-spread acquisition geometry with 50 sources and 246 receivers placed at 25m below the water surface. The sources are evenly distributed in the horizontal direction from 50m to 6175m with a spacing of 850m, while the receivers are also evenly spaced from 50m to 6175m with a spacing of 25m. The source time function is a high-pass filtered Ricker wavelet with a dominant frequency of 4.5Hz (Figure~\ref{fig:Source_time_function}(a)). Figure~\ref{fig:Source_time_function}(b) shows that the frequency contents below 2Hz are attenuated since they are hard to obtain in real situations. We create synthetic datasets by numerically solving Equation~(\ref{eq:ConstantAcousticWaveEq}) using a staggered-grid finite-difference solver with 4th-order accuracy in space and 2nd-order accuracy in time. One example of shot gather from data without noise is shown in Figure~\ref{fig:Filtered_data}(a). We also create a noisy dataset by adding in white Gaussian noise such that each shot gather has a signal-to-noise ratio (SNR) equaling to 10 (Figure~\ref{fig:Filtered_data}(b)).

We use all 50 sources in the regular FWI. To save computation time, in TV and NMAS cases we use cyclic shooting as in \citet{ha2013efficient}, with 10 sources for each outer loop iteration. To keep the aperture the same, we retain source No. 1 and source No. 50 in all outer loop iterations, and only change the 8 sources used in between.

In inversions with regularizations, we choose weighting factors $\rho$ and $\beta$ by setting $R_{\rho} = 2\times 10^{-3}$ and $R_{\beta} = 2\times 10^{-3}$ (see section \ref{sec:algorithm}). We handle the FWI with TV regularization using the same framework as NMAS, since for TV we can just set the window size to be the size of the whole model and let the dictionary be an identity matrix. In the NMAS case we soft-threshold the training patches by the ratio of $\beta/\rho$, and we set the number of classes to be 36. The total number of outer loop iterations is 13, and the inner loop iteration number starts from 30 and increases by 10 per each outer loop.

Inversion results are shown in Figure \ref{fig:BP_results}. The top two panels are the true and initial velocity models respectively. The group of Figures \ref{fig:BP_results}(c), (d) and (e) on the left show inversion results without noise in the data, and the group of Figures \ref{fig:BP_results}(f), (g) and (h) on the right show results with added random noise in the data for $\text{SNR} = 10$. In each group, we show the results of FWI without regularization, with TV regularization, and with the proposed NMAS regularization, respectively.

In the noise-free case, the traditional FWI without regularization reconstructs the general features of the model, and the top part of the salt body is resolved reasonably well. However, the traditional inversion result has strong artifacts. Also, the salt boundaries are not accurately delineated and the sub-salt area is poorly resolved because of poor illumination (see for example the circled area in Figure~\ref{fig:BP_results}(c)). FWI with TV produces a greatly improved result when compared to FWI without regularization, but the features are blocky and are not realistic in the geological sense (see the circled area in Figure~\ref{fig:BP_results}(d)). The inversion result with the NMAS method further improves the result from TV, with salt body boundaries and sub-salt velocity structures more accurately resolved. We see a similar situation with the presence of noise in the data. The FWI result without regularization has more artifacts. The FWI result with TV regularization are blockier, and even the boundary at the bottom left part of the model is smoothed out (see the circled area in Figure~\ref{fig:BP_results}(g)). One reason for this we suspect is that $\rho$ and $\beta$ change as the norm of data changes. The NMAS result has the best resolved salt-body and sub-salt structures, though it has mild noise.

Our analyses above are qualitative. Let us now examine our results quantitatively. Since in this synthetic test we know the exact true velocity model, we can numerically compare our inverted models with this exact model to quantify inversion quality. Here we use two measures to compare models, one is the normalized model mean squared error (MSE), and the other one is the structural similarity index (SSIM) \citep{wang2004image}, which is an improvement on both MSE and peak signal-to-noise ratio (PSNR). The SSIM value varies from -1 to 1, with a higher value interpreted as the two images being more similar, and 1 means they are identical.

\begin{figure}
\begin{center}
\includegraphics[width=\textwidth]{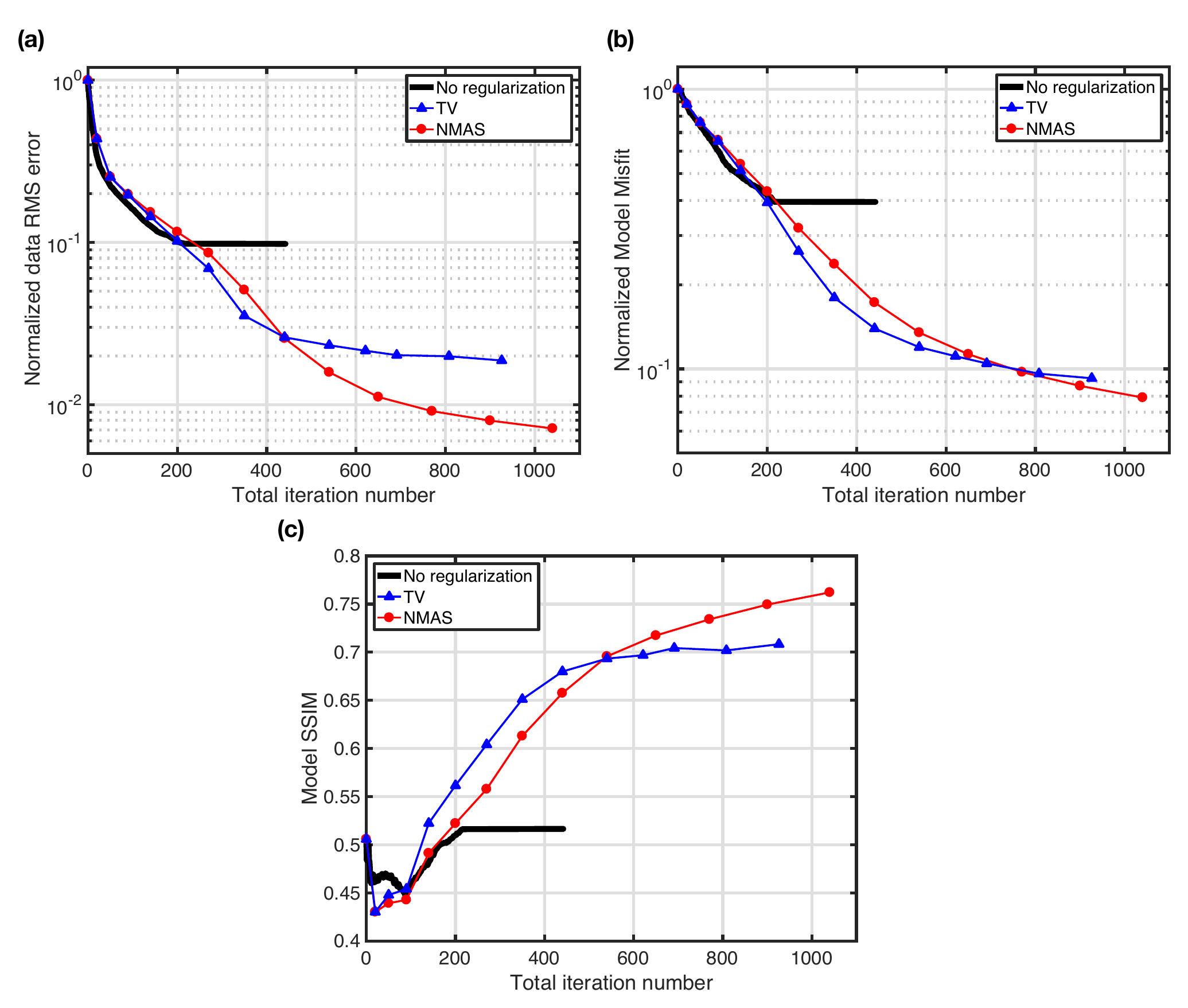}
\end{center}\vspace{-0.5cm}
\caption{Misfit curves for BP 2004 inversion results from three different methods. (a). Normalized data root-mean-square curve; (b). Model mean squared error curve; (c). Model SSIM curve. Note the first two curves are in log scales in the vertical axis.}
\label{fig:Misfits}
\end{figure}

For each inversion result in the noise-free experiment, we plot three curves: a normalized data root-mean-square misfit (RMSE) curve, a normalized model mean squared error curve, and a SSIM curve, which are shown in Figure \ref{fig:Misfits}. For TV regularization and NMAS regularization, we only plot the values at the end of each outer loop.

The data misfit curve of FWI without regularization converges around iteration 210, where inversion seems to get trapped in local minima, and stops at iteration 444 because the stopping criteria of the L-BFGS solver are met. The data misfit curve of TV regularization further reduces the data misfit, while NMAS is able to reduce it the most, which means that the model inverted by FWI with NMAS regularization is able to explain the observed data the best (Figure~\ref{fig:Misfits}(a)). However, this does not tell the whole story. Since the solutions are non-unique, there are multiple models that can explain the data equally well. Therefore, we compare the normalized model MSE and model SSIM. We find that FWI with NMAS regularization can also reduce model MSE the most and has the highest SSIM (Figure~\ref{fig:Misfits}(b) and (c)). These results quantitatively verify that our NMAS method for this experiment achieves the best result. Note the total iteration number of TV is smaller than that of NMAS. Although we gradually increase the maximum number of iterations allowed for inner loops, TV seems not be able to use all of those inner iterations during later outer loops because the L-BFGS solver meets its stopping criteria early, causing an early termination.

\subsection{Sensitivity Analyses on Parameters}
\begin{figure}
\begin{center}
\includegraphics[width=16cm]{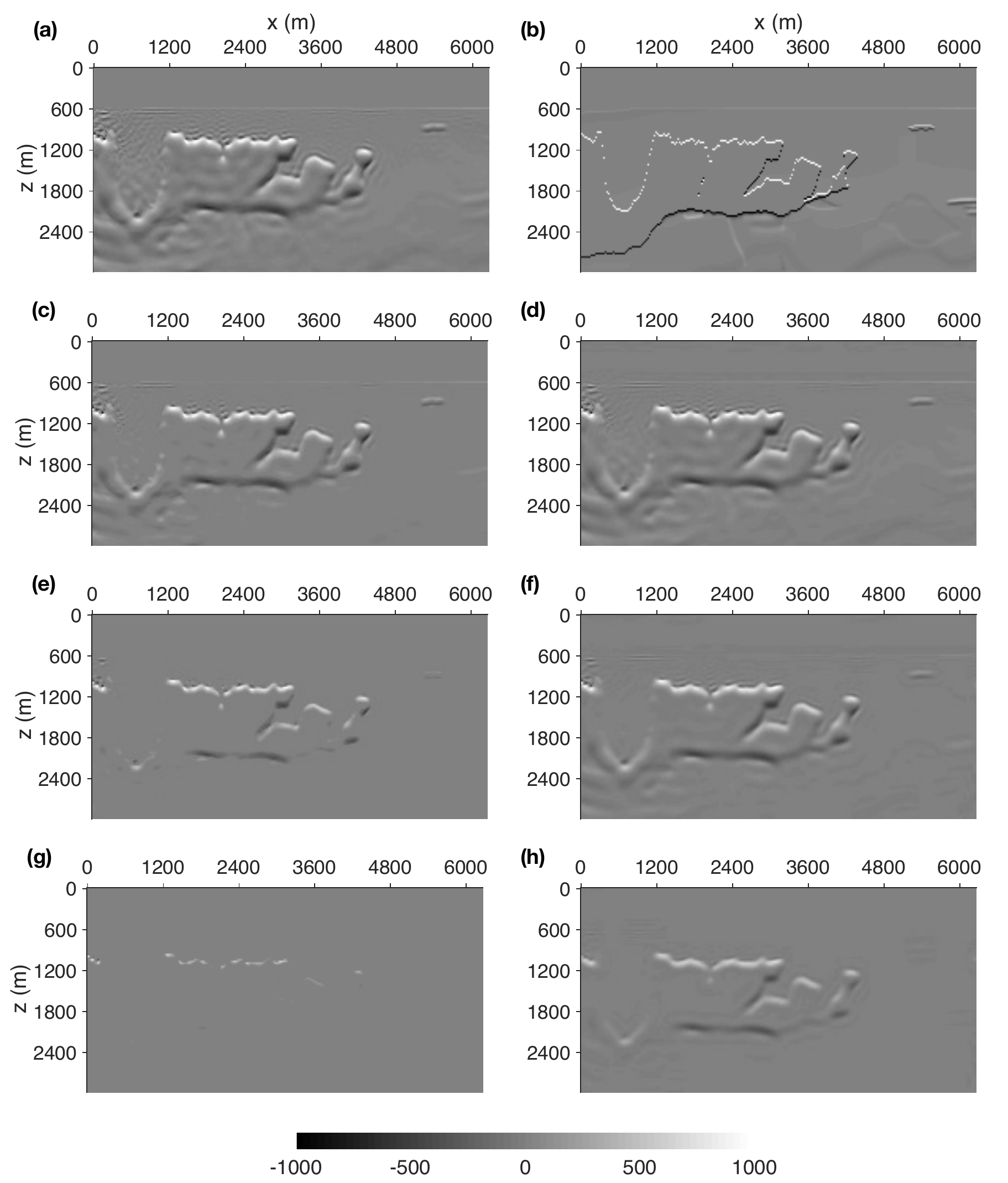} \\[-0.7cm]
\end{center}\vspace{-0.15cm}
\caption{The sensitivity of the sparse-approximated model derivative to the thresholding parameter for TV and NMAS. (a). The provided vertical model derivative from an intermediate inverted model, which has artifacts; (b). The vertical model derivative of the true model. (c), (e), and (g) are sparse-approximated vertical model derivatives in TV with the thresholding parameter as 50, 200, and 600, respectively; (d), (f), and (h) are sparse-approximated vertical model derivatives in NMAS with the thresholding parameter as 50, 200, and 600, respectively. The derivatives in TV are more sensitive to the thresholding parameter.}
\label{fig:gradient_sensitivity}
\end{figure}

In the preliminary results shown above, NMAS does give us better results than TV, but this does not sufficiently prove the superiority of NMAS over TV, and that is not our purpose. In fact, we admit that FWI with TV regularization can also generate results of high quality close to NMAS, if we pay extra effort to fine tuning the parameters. However, we observe that NMAS results are more robust to parameter tuning than TV. In other words, if we change the weighting parameters by the same amount, TV results can have large changes while the NMAS results do not. Since the inversion results are obtained with the guidance of the sparse-approximated model derivatives, we examine their sensitivity to those parameter changes. According to Equation (\ref{eq: alpha_update3}), the sparse-approximation coefficients are determined by the ratio $\beta/\rho$, so in this sensitivity analysis we vary this ratio from low value to high value. As shown in Figure \ref{fig:gradient_sensitivity}, as we increase $\beta/\rho$ from 50 to 200, the sparse-approximated model derivative in TV has fewer and fewer artifacts while at the same time has sharper and sharper boundaries around the salt body. However, geological structures also gradually disappear in this process. In Figure \ref{fig:gradient_sensitivity}(c) the structure at the lower-right corner disappears; in Figure \ref{fig:gradient_sensitivity}(e) only a rim of salt body is available; in Figure \ref{fig:gradient_sensitivity}(g), only the top of salt body exists. Therefore, if we use the model derivative obtained with a too low $\beta/\rho$ ratio to guide the inversion, we will have results exhibiting high level of artifacts, but if we use the model derivative with a too high $\beta/\rho$ ratio, we will have results with overly sharpened edges and overly smoothed structures. On the contrary, we see that in the same process, the sparse-approximated model derivatives from NMAS always have the structure of the salt body even at the highest ratio. Also, structures below the salt body are better preserved. We infer that this is because dictionary atoms are like geological structures and have geological meanings. Under thresholding only a small number of those atoms will be used for approximation, but those small number of atoms capture the most important geological information. So even when the ratio changes wildly, the major information is kept, whence a better preservation of structures and reduced artifacts. 

In the NMAS method, the window size controls how much patterns can vary within one patch. In addition, the number of dictionaries or patch groups controls how many patches are similar to each other. If we have a small number of patch groups, for example, the whole set of patches would be divided into only a few groups, which imposes more constraint on the inversion. Therefore, finally we test the sensitivity of the inversion results to those two parameters, i.e., $n$ and $s$. 

As a reminder, we recap the parameters used in the base case. The ratios for determining $\rho$ and $\beta$ are both set as $R_{\rho} = 2\times 10^{-3}$ and $R_{\beta} = 2\times 10^{-3}$, the window size is 8 grid points, and the number of classes is 36.

First we fix all other parameters, and use window size of 4, 8, and 16. The SSIM of the inversion results are 0.7616, 0.7621, and 0.7553, respectively. Then we fix all other parameters, and use number of classes of 18, 36, and 72. The SSIM of the results are 0.7583, 0.7621, and 0.7572. The SSIM slightly changes as the size of window and the number of classes change. However, the values are stable, meaning that the inversion results are robust with respect to those two parameters.


\subsection{Modified BP 2004 Model with Higher-frequency Data}
\begin{figure}
\begin{center}
\includegraphics[width=0.8\textwidth]{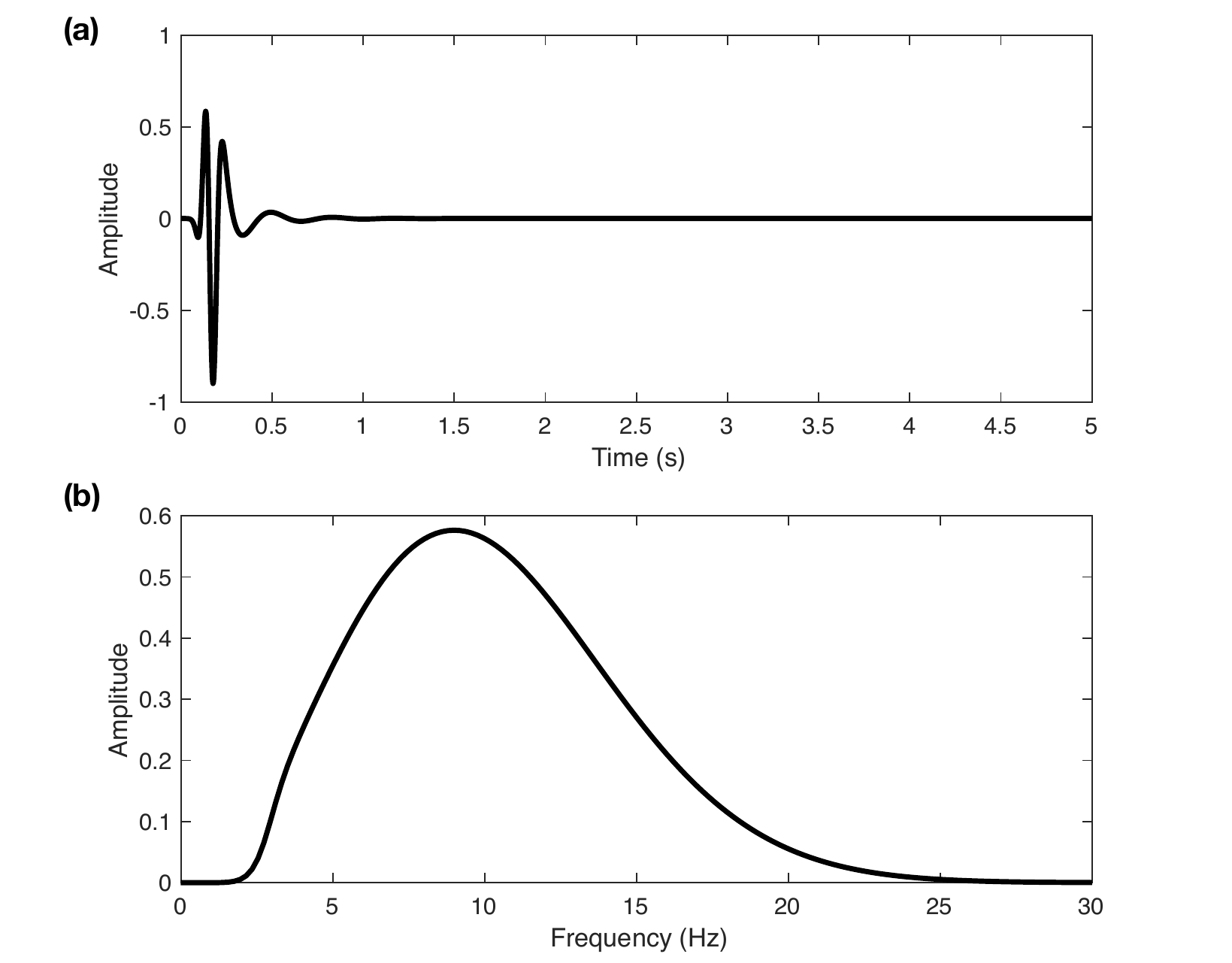}
\end{center}\vspace{-0.5cm}
\caption{Source-time function with a dominant frequency of 9Hz and its spectrum. (a). The source-time function; (b). The spectrum. The frequency contents below 2Hz are attenuated.}
\label{fig:Source_time_function_9Hz}
\end{figure}

\begin{figure}
\begin{center}
\includegraphics[width=\textwidth]{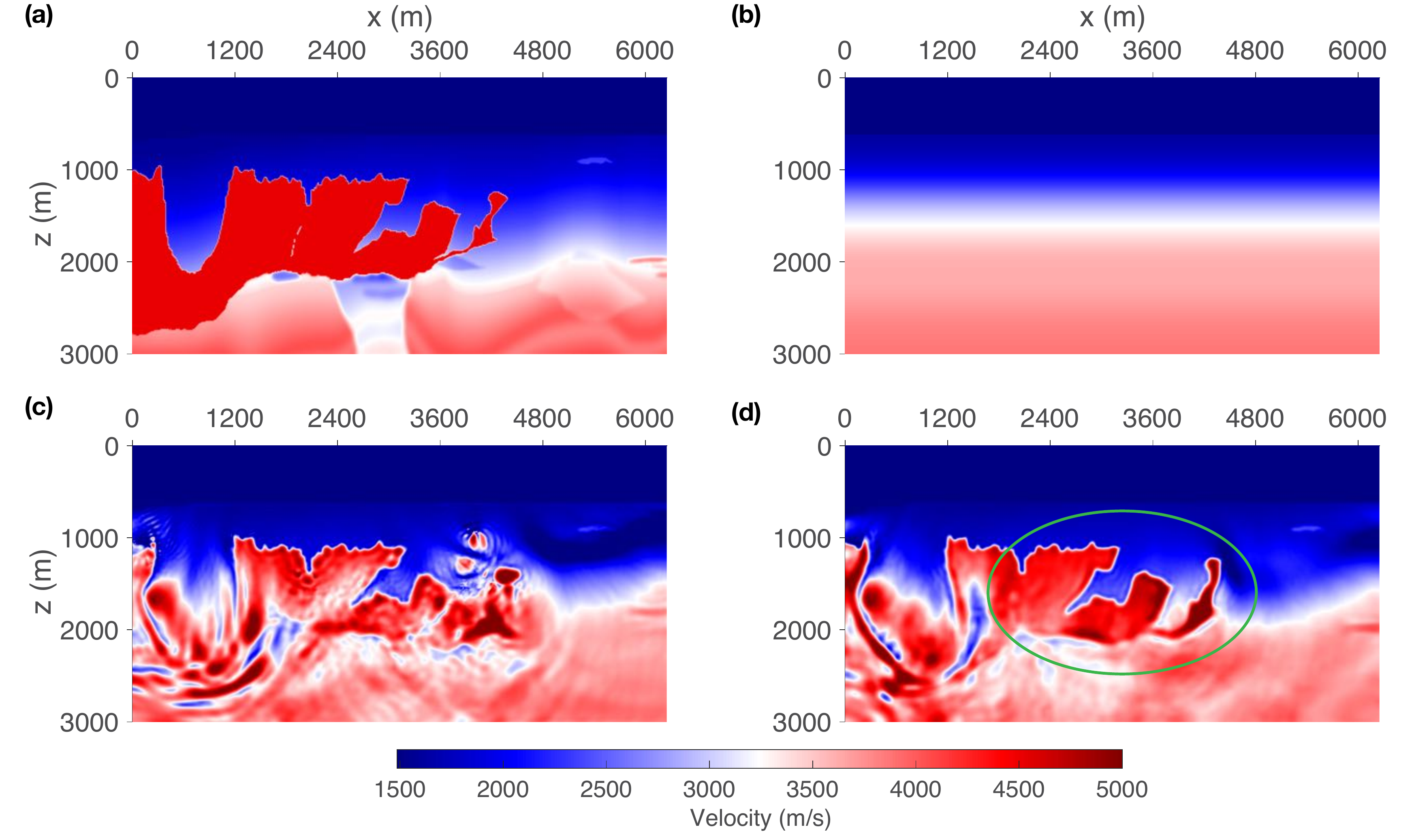}
\end{center}\vspace{-0.5cm}
\caption{The FWI inversion results of the BP 2004 model with higher-frequency data. (a). The true velocity model; (b). The 1-D initial model; (c). Traditional FWI result without regularization (SSIM=0.3215); (d) NMAS-FWI result (SSIM=0.4150). Both the FWI-without-regularization result and the NMAS-FWI result are trapped in local minima, but part of the NMAS-FWI result is more accurately resolved and has fewer artifacts (see the circled area).}
\label{fig:BP_results_9Hz}
\end{figure}

As mentioned in the introduction, if we start directly with data containing higher-frequency contents, FWI is more susceptible to the cycle-skipping problem. To test the performance of our NMAS method in this scenario, we double the dominant frequency of the Ricker wavelet to 9Hz. The highest frequency extends beyond 20Hz, and we filter out frequency contents below 2Hz as before (Figure \ref{fig:Source_time_function_9Hz}). We keep other experiment settings the same as the previous experiment. The inversion results are shown in Figure \ref{fig:BP_results_9Hz}. 

We see that the results from FWI without regularization and from our proposed NMAS-FWI are both trapped in local minima. The subsalt area is poorly resolved, and both results have strong artifacts. This suggests that our NMAS regularization cannot completely overcome the problem of cycle-skipping. However, compared with the traditional FWI result, the central part of our NMAS-FWI result is more accurate and has a lower level of artifacts. This fact suggests that the NMAS regularization might mildly reduce the non-convexity of the objective function, and also promotes coherent structures in the inversion result.

\subsection{Smoothed Marmousi Model}

\begin{figure}
\begin{center}
\includegraphics[width=\textwidth]{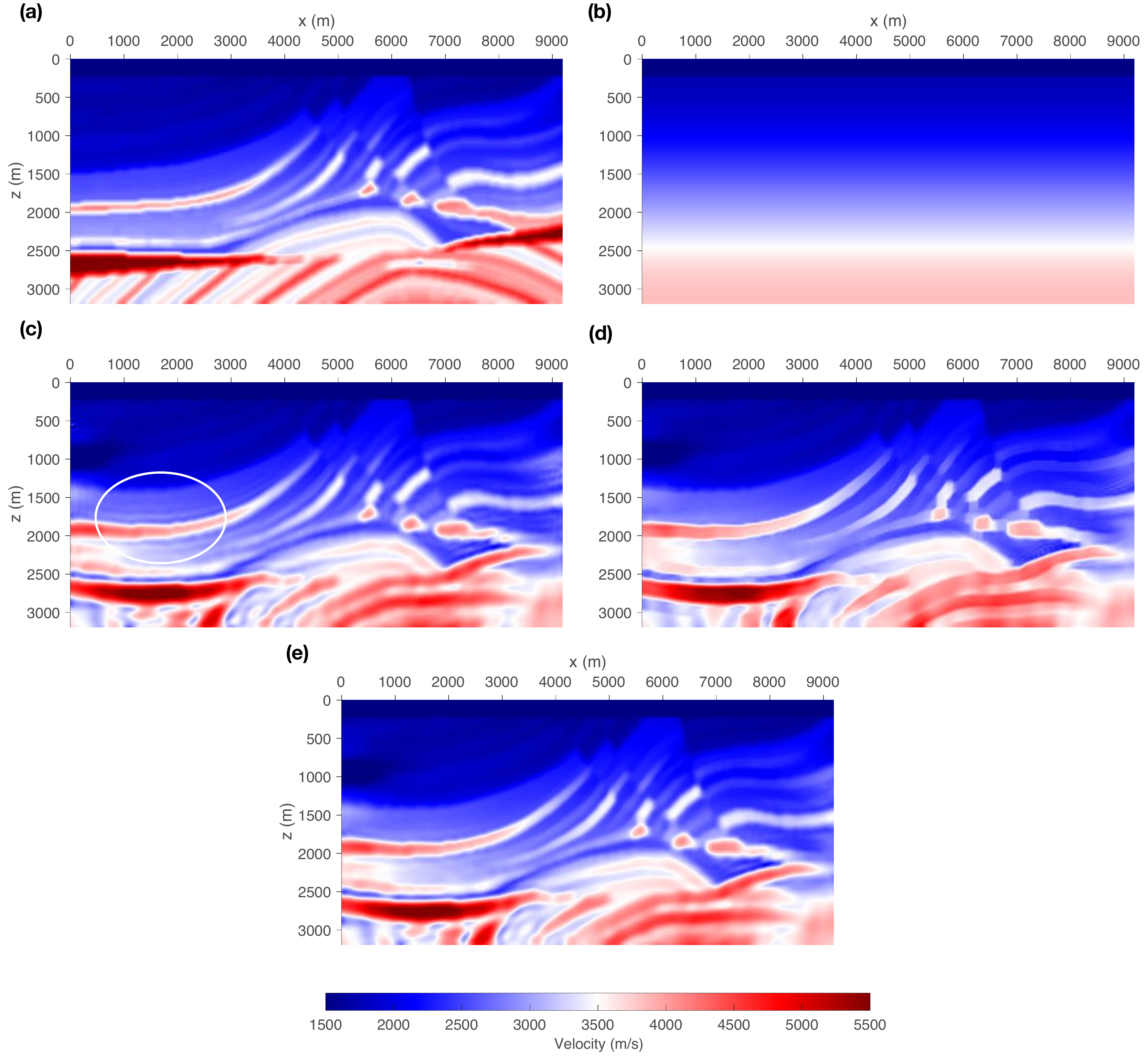}
\end{center}\vspace{-0.5cm}
\caption{The FWI inversion results of the smoothed Marmousi model. (a). The true velocity model; (b). The 1-D initial model; (c). The inversion result without regularization; (d) The inversion result with TV regularization; (e). The inversion result with NMAS regularization.}
\label{fig:Mar_results}
\end{figure}

\begin{figure}
\begin{center}
\includegraphics[width=\textwidth]{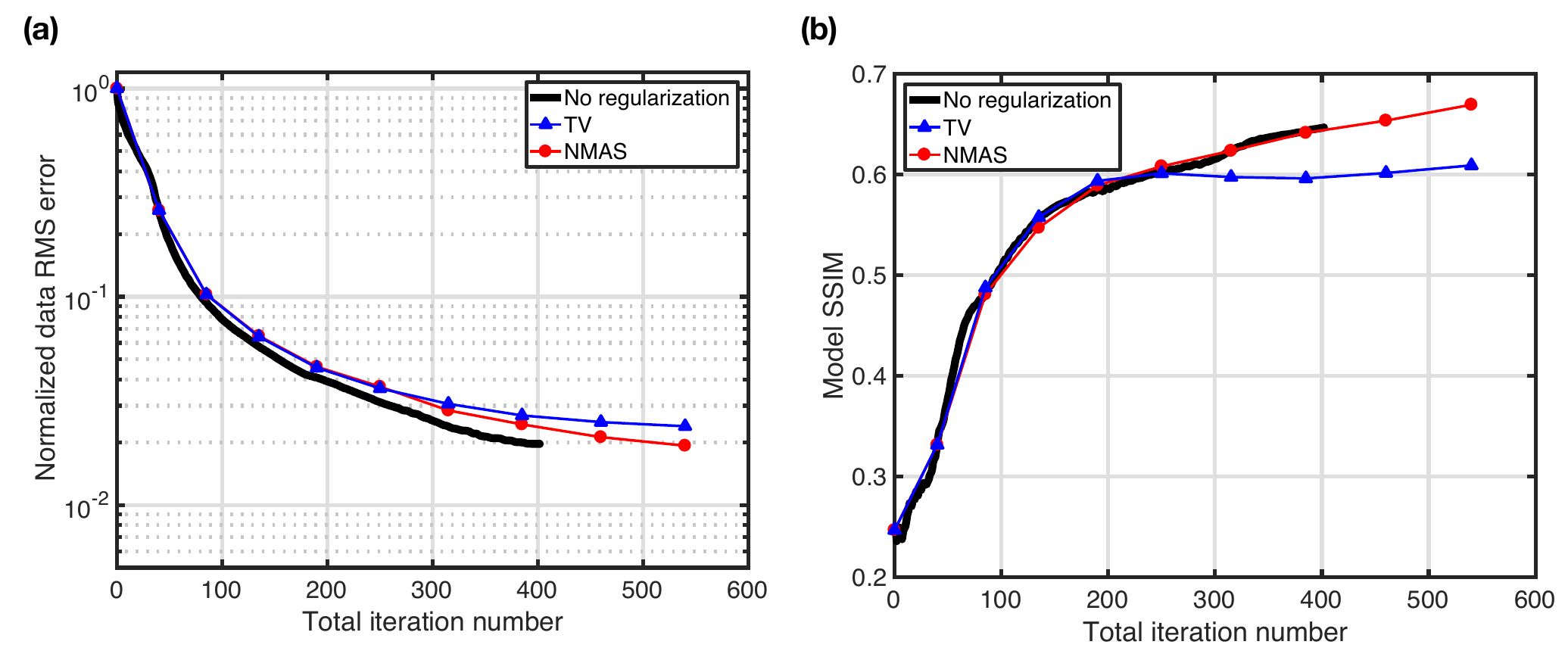}
\end{center}\vspace{-0.5cm}
\caption{Misfit curves of Marmousi inversion results from three different methods. (a). Normalized data root-mean-square curve; (b). Model SSIM curve.}
\label{fig:Misfits_Mar}
\end{figure}

The previous inversion experiments are based on a model with high velocity contrast and few layer structures. In this experiment, we test our method on a slightly smoothed Marmousi model \citep{versteeg1994marmousi}, which has weaker velocity contrast and more reflectors.

The model is about 9.2km wide and 3.2km deep, with a 200m layer of water on top. The grid spacing is 24m. We create the true velocity model by smoothing the original Marmousi model using a $5\times5$ mean filter (Figure \ref{fig:Mar_results}(a)). We then create the 1-D initial model by first smoothing the slowness of the true model using a $101\times 101$ Gaussian filter with a standard deviation of 15 grid points, and then averaging the velocity along the horizontal axis (Figure \ref{fig:Mar_results}(b)).

We use a fix-spread acquisition geometry with 48 sources and 379 receivers placed at 24m deep in the water. The source-time function is the same as in the first BP 2004 model experiment, whose dominant frequency is 4.5Hz with frequency contents below 2Hz attenuated (Figure \ref{fig:Source_time_function}).

As shown in Figure \ref{fig:Mar_results}(c), traditional FWI without regularization produces a good inversion result, where the detailed features are accurately reconstructed, at least in places above 2500m, where the fold is relatively high. The presence of abundant diving waves in the data due to the shallow depth of water may contribute to this success. Similar to the no-regularization result, our NMAS-FWI result is also of high resolution and accuracy (Figure \ref{fig:Mar_results}(e)). If we examine the results carefully, at certain places, in the circled area for example, the no-regularization result has spurious thin layers, while our method reduces such artifacts. However, our method can neither accurately reconstruct the complex layer structures below 2500m. Perhaps structures at shallower depth do not provide enough similarity constraint for the set of parallel dipping layers and the antiform. This could be one limitation of our method.

The TV method reconstructs layers with very sharp edges and nearly constant velocities, which are different from true structures. It confirms that the blockiness-promoting prior behind TV may not represent real geology. Our method has an advantage over TV in this scenario.

Numerically, our NMAS method reduces data misfit the most, slightly better than that of the traditional FWI. It also increases the model SSIM the most among the three methods (Figure \ref{fig:Misfits_Mar}).

Our method stops at the 540th iteration while FWI without regularization stops at the 403th iteration. We do not use as many number of iterations in the inner loop as the previous BP 2004 tests, since we know that in this Marmousi experiment it is not as hard to converge. We will discuss the extra cost to implement nonlocal similarity in the following section.

\section{Discussion}


The computational cost of our NMAS-FWI is higher than traditional FWI without regularization, mainly because of the nested loops and the extra operations related to nonlocal similarity regularization: patch extraction, HoG feature descriptor calculation, clustering, orthogonal dictionary learning, and sparse approximation calculation. 

For the nested loops, in our first BP 2004 experiment the number of outer loops and inner loops are relatively high, which could be a drawback of our method. It seems that in this case our NMAS regularization mitigates the local minima problem and FWI converges to a good solution. However, the convergence is slow, perhaps due to using the poor 1-D starting model, which is too far from the true model, and the limited of number of shots (we only use a subset of all shots in each inner loop). In the Marmousi experiment, the model is easier to resolve and we use all shots in the inversion, so the necessary number of inner loops and outer loops are reduced. When dealing with field data, we do not need to use as many inner loop iterations as the BP 2004 experiments, since field data are usually very noisy and it is not recommended to iterate towards full convergence, trying to fit data to the best. Also, we may explore optimization solvers with faster convergence performances in the inner loop and use source encoding strategies to speed up computation \citep{krebs2009fast}.

As for the extra operations, we calculate their time complexity in the big-$\mathcal{O}$ notation. Let $M$ denote the number of patches, and $N=n^2$ the number of points in one patch. The time complexity of patch extraction and HoG feature descriptor computation is $\mathcal{O}(MN)$, and the complexity of clustering in one iteration is $\mathcal{O}(MKd)$, where $K$ here is the number of clusters, and $d$ is the dimension of feature descriptors. The complexity of one iteration in dictionary learning is $\mathcal{O}(MN^2 + N^3)$ \citep{bao2013fast}. Also, the cost of sparse approximation computation is $\mathcal{O}(MN^2)$ since mainly matrix multiplication and soft-thresholding are involved. Therefore, the whole time complexity of the extra operations may be dominated by clustering and dictionary learning. Empirically the clustering can finish quickly as $K$ and $d$ are usually small numbers, which are 36 and 9 respectively in our examples. Also, $N$ is usually a small number, such as 64 for a $8\times 8$ patch, making orthogonal dictionary learning fast. Besides, the number $M$ in each class is small, and the multi-class dictionary learning can be easily parallelized by sending dictionary learning for different classes to different CPU cores, further reducing the computation time. In our experience, on an Intel Xeon E5-2640 v3 2.6GHz 8-core processor, it takes around 23 seconds to finish all the extra operations for 32766 $8\times 8$ patches in 36 classes from the vertical model derivative.

There are two limitations in our method. First, as demonstrated in the Marmousi model, if structures in poorly constrained area are not well represented in other places with better constraint, our nonlocal similarity prior cannot effectively recover them. Second, our method is not guaranteed to solve the local minima problem but to some degree mitigate it. Therefore, in practice our method should be combined with other strategies that are specifically developed for this purpose. Examples include the frequency continuation method \citep{pratt1999seismic} or the multiscale method \citep{bunks1995multiscale}, extension methods \citep{symes1991velocity,biondi2014simultaneous,warner2016adaptive,gao2014new,van2013mitigating}, and using objective functions less prone to local minima \citep{luo1991wave,metivier2016measuring,yang2018application}. As a regularization method, NMAS should work with all those methods mentioned above.

Elastic effects are inevitably present in the data. In an acoustic inversion, shear waves or converted waves cannot be predicted by an acoustic forward modeling engine. Also, the AVO effects are calculated wrong. These would cause coherent artifacts in inversion results. It needs further investigations to know whether our NMAS regularization can help acoustic FWI deal with this situation. On the other hand, however, we can switch to elastic NMAS-FWI if necessary since the theory is almost the same.

Our proposed method can be viewed as a generalization of TV regularization, but differs in how the sparse-approximated derivatives are generated. In TV, the model derivatives are generate by directly soft-thresholding the original derivatives, without regard to geological patterns. In our method, however, we soft-threshold the sparse-coefficients under dictionaries of geological patterns. This makes it more effective in getting rid of noise and artifacts while at the same time preserving geological structures. This also makes our method less sensitive to parameter tuning. However, since TV itself is a powerful method for suppressing noise, we may combine TV and our method together in the same framework. That is, in early iterations, we use TV since the model derivatives have few geological patterns and are full of artifacts. In later iterations, however, as geological patterns appear, we can switch to NMAS. Also, we can add soft-thresholding in the patch generation step to help NMAS reduce artifacts.

In NMAS method, multi-class orthogonal dictionaries are used to implement the nonlocal similarity prior. In the process of generating training patches for dictionary learning, we take multi-resolution into account, as the patches are extracted under translation, scaling and rotation operations. 

The major goal of our work is to introduce nonlocal similarity in FWI, so we only use internal information. That is, we perform inversion in a bootstrapping way that begins without outside information. In practical applications, external information may be available and valuable. For example, we may have seismic migration images or inversion results of well studied areas that are of similar geological settings as the one to be investigated. Or, we may have plausible models constructed by geologists based on other geological or geophysical information. In these scenarios, we may want to combine external information and internal nonlocal similarity to further speed up convergence and obtain a better solution. We leave this strategy for future studies.

Our method fits in the framework of ADMM algorithm; thus, some techniques for parameter tuning can be adopted. For example, in \citet{boyd2011distributed}, the parameter $\rho$ can be chosen adaptively as iterations proceed. This strategy may also speed up convergence.

Our proposed method works in the model-derivative domain, primarily because model derivatives are naturally sparse and thus it is easy to find patches of similar features. It is also possible to extend our method to other domains, such as the wavelet domain. In fact, in this paper we provide a general framework to introduce nonlocal similarity with dictionaries, which can be applied to other domains easily.

Currently we only consider 2-D applications. In the future, we may explore applying this method to 3-D inversion cases. In that setting, the dimension of the 3-D dictionaries increases dramatically, so the computation and memory cost of dictionary learning and the averaging of 3D patches increases dramatically. One possible strategy is to only use 2-D patches in 3-D inversions. This poses a future research direction.

Finally, we believe that our proposed NMAS regularization can be applied to other seismic inversion methods, such as ray-based tomography.

\section{Conclusions}
In this paper, we have presented a regularization method (NMAS) for FWI that incorporates a nonlocal similarity prior. It is closely related to image processing techniques and convolutional neural networks. Implemented with sparsity-promoting multi-class dictionary learning, our method reduces the degrees of freedom in model parameters and increases the coherency and resolution in the inversion results. It may also mitigate the problem of local minima in FWI. Compared with TV regularization, our method is less sensitive to parameter tuning and produces results that are more geologically meaningful. We have also pointed out limitations of our proposed method, and provided suggestions for practical use. As a regularization technique, our method can be used together with other FWI formulations to further improve inversion results.

\section*{Acknowledgements}
We thank Stewart A. Levin for his comments and suggestions that greatly improved the quality of this manuscript. We thank Biondo L. Biondi for helpful discussions. We greatly appreciate the constructive and detailed comments and suggestions from David Al-Attar and an anonymous reviewer. The first author acknowledges financial support from Stanford Wave Physics (SWP) Laboratory.

\newpage
\bibliography{NLGD_ref.bib}

\appendix

\section{Total Variation Regularization}
\label{sec:TV}

\begin{figure*}
\begin{center}
\includegraphics[width=\textwidth]{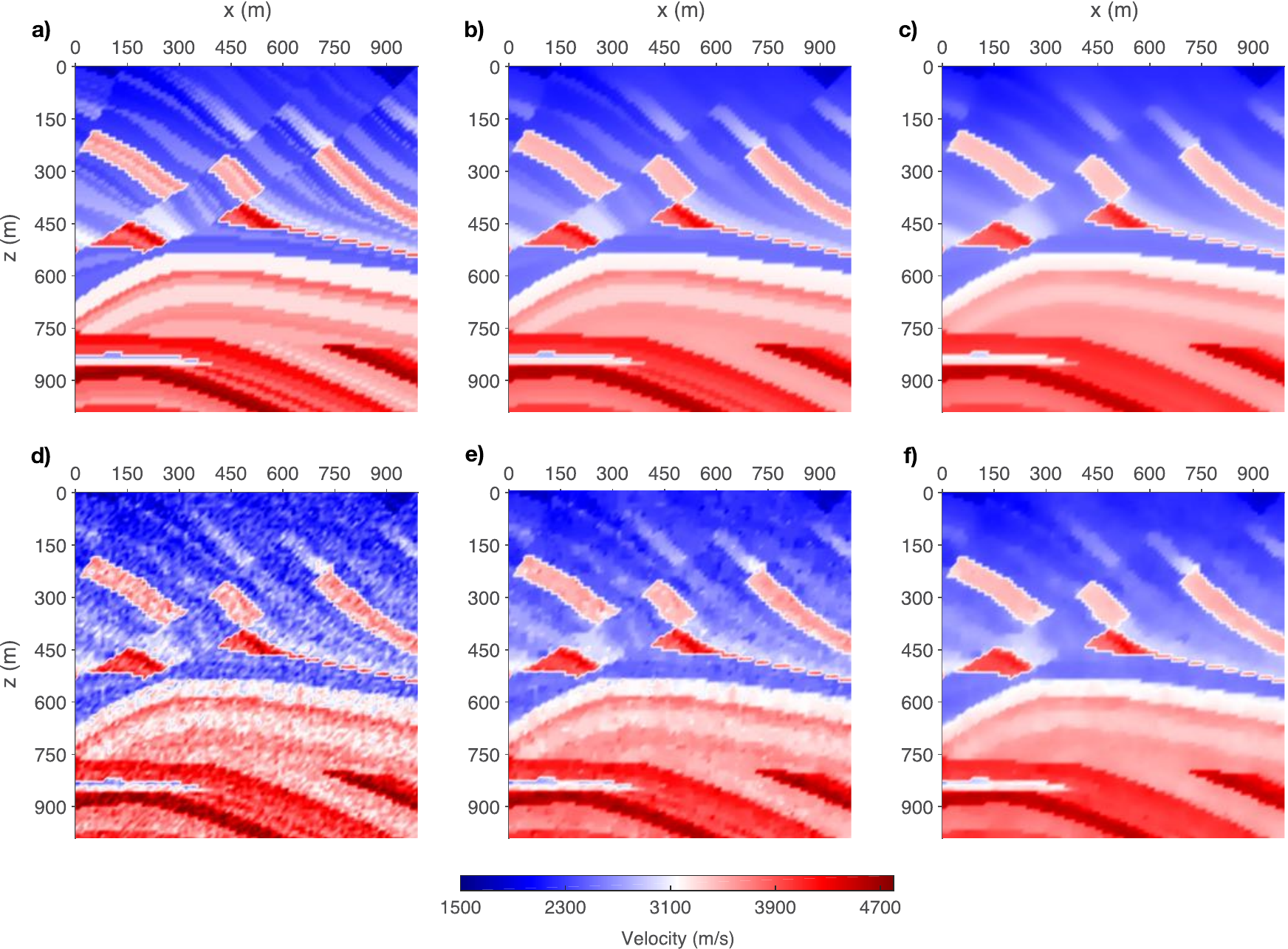}
\end{center}\vspace{-0.5cm}
\caption{Effects of total variation on velocity model approximation. (a). Noise-free velocity model; (b). Velocity model reconstructed from noise-free velocity model with a small $\lambda$; (c). Velocity model reconstructed from noise-free velocity model with a large $\lambda$; (d). Velocity model with noise; (e). Velocity model reconstructed from noisy velocity model with a small $\lambda$; (f). Velocity model reconstructed from noisy velocity model with a large $\lambda$.}
\label{fig:Mar_tv}
\end{figure*}

To find high resolution models, blocky-model promoting regularizations such as total variation \citep{rudin1992nonlinear} are adopted. Their prominent characteristic is to preserve eminent edges or sharp discontinuities, while at the same time smooth out features with small gradients, such as weak or mild noise. In fact, the mathematical form of a (discrete) TV term is
\begin{equation}
\label{eq:tv_def}
	\|\mathbf{m}\|_{\text{TV}} = \sum_{i,j} \left\Vert \begin{pmatrix} \nabla_\mathrm{v} \mathbf{m}_{i,j} \\\nabla_\mathrm{h} \mathbf{m}_{i,j}\end{pmatrix} \right\Vert = \sum_{i,j} \left\Vert \begin{pmatrix} \mathbf{m}_{i+1,j}  - \mathbf{m}_{i,j} \\  \mathbf{m}_{i,j+1} - \mathbf{m}_{i,j}\end{pmatrix} \right\Vert,
\end{equation}
where $\nabla_\mathrm{v}$ and $\nabla_\mathrm{h}$ are finite-difference operators along the vertical and horizontal directions, respectively. The norm in Equation (\ref{eq:tv_def}) can be either $\ell_2$-norm or $\ell_1$-norm, which respectively correspond to isotropic and anisotropic TV terms. Specifically, the anisotropic TV of $\mathbf{m}$ is 
\begin{equation}
\label{eq:tv_anis}
\begin{split}
	\|\mathbf{m}\|_{\text{TV}, 1}  & = \sum_{i,j} \left\Vert\mathbf{m}_{i+1,j} - \mathbf{m}_{i,j}\right\Vert_1 + \left\Vert\mathbf{m}_{i,j+1} - \mathbf{m}_{i,j}\right\Vert_1 \\
	& = \|\nabla_\mathrm{v} \mathbf{m}\|_1 + \|\nabla_\mathrm{h} \mathbf{m}\|_1,
\end{split}
\end{equation}
which is the sum of $\ell_1$-norm of the vertical and horizontal gradients/derivatives of the model. Note that in the second row in Equation (\ref{eq:tv_anis}), the terms $\nabla_\mathrm{v}\mathbf{m}$ and $\nabla_\mathrm{v}\mathbf{m}$ should be understood as the vectorized gradient images. Since the $\ell_1$-norm is a measure of sparseness, the TV regularization enforces sparsity on the model gradients. It penalizes small gradient jumps whereas preserves large gradients. This corresponds to the prior knowledge that the model is blocky or piecewise constant. To better visualize this, consider the following problem
\begin{equation}
	\argmin_\mathbf{m} \|\mathbf{m} - \mathbf{u}\|^2_2 + \lambda \|\mathbf{m}\|_\mathrm{TV},
\end{equation}
where $\mathbf{u}$ is known model, and we want to find a model close to it with small total variation controlled by trade-off parameter $\lambda$. Solved by the splitting method proposed by \citep{wang2008new}, this problem has the following results shown in Figure \ref{fig:Mar_tv}. First, we have a noise-free velocity model from the Marmousi model shown in Figure \ref{fig:Mar_tv}(a). As we increase the TV weighting factor $\lambda$, the reconstructed velocity model still has the structural boundaries but the fine details within those boundaries are gradually smoothed out (Figure \ref{fig:Mar_tv}(b) and (c)). This is because TV regularization requires the gradient be sparse, and only boundaries with strong values jumps are kept. If we add noise into model and redo this experiment, we see that in addition to the phenomenon that we described, the noise is more and more reduced as $\lambda$ increases. This tells us that if the noise or artifacts are not strong enough to cover the main structures in the velocity model, TV regularization can effectively attenuate them since the gradient of noises are relatively small.

TV regularization also has some limitations. It tends to create blocky models, which is not always realistic. Also it is an analytic regularization technique that is not able to adapt to geological structures. This sometimes makes it hard to effectively preserve those structures while getting rid of noises and artifacts.

\section{Alternating Direction Method of Multipliers}
\label{sec:admm}
Here we first discuss the Augmented Lagrangian method, which is usually used to solve equality-constrained optimization problems:
\begin{equation}
\begin{array}{ll}
	\mbox{minimize} & E(x) \\
	\mbox{subject to} & Ax = b.
\end{array}
\end{equation}
The augmented Lagrangian is 
\begin{equation}
	L_{\rho}(x, \lambda) = E(x) + \lambda^T(Ax-b) + (\rho/2)\|Ax-b\|_2^2, 
\end{equation}
where $\lambda$ is the Lagrange multiplier, and $\rho$ is the penalty parameter \citep{boyd2011distributed}.
It can be solved by the following iterative scheme:
\begin{equation}
\begin{split}
	& x^{k+1} := \argmin_x L_{\rho}(x, \lambda^k) \\
	& \lambda^{k+1} := \lambda^k + \rho(Ax^{k+1} - b).
\end{split}
\end{equation}

If the objective function can be decomposed into multiple parts (two in this case without loss of generality), for example,
\begin{equation}
\label{eq:ADMM_objective}
\begin{array}{ll}
	\mbox{minimize} & f(x) + g(z) \\
	\mbox{subject to } & Ax + Bz = c.
\end{array}
\end{equation}
Then we have the augmented Lagrange as
\begin{equation}
	L_{\rho}(x, z, \lambda) = f(x) + g(z) + \lambda^T(Ax+Bz-c) + (\rho/2)\|Ax+Bz-c\|_2^2.
\end{equation}
Accordingly, the iterative scheme is
\begin{equation}
\label{eq:ADMM_iterative}
\begin{array}{ll}
	& x^{k+1} := \argmin_x L_{\rho}(x, z^k, \lambda^k) \\
	& z^{k+1} := \argmin_z L_{\rho}(x^{k+1}, z, \lambda^k) \\
	& \lambda^{k+1} := \lambda^k {}+ \rho(Ax^{k+1} + Bz^{k+1} - c)
\end{array}
\end{equation}

This is particularly useful in solving regularization problems. Consider an unconstrained problem:
\begin{equation}
	\mbox{minimize} \quad f(x) + g(Ax).
\end{equation}
Sometimes it is difficult to solve the original problem as a whole, but it is much easier to minimize $f(x)$ and $g(x)$ separately. One example is that the regularization term is the $\ell_1$ constraint, which is not differentiable. In this case, we may change the variable in one misfit function, add one constraint to force the new variable to be equal to the original one, and have 
\begin{equation}
\begin{array}{ll}
\mbox{minimize} & f(x) + g(z) \\
\mbox{subject to} & Ax - z = 0, 
\end{array}
\end{equation}
which becomes the same formulation as in Problem (\ref{eq:ADMM_objective}), and can be solved using the iterative scheme (\ref{eq:ADMM_iterative}). This is the so-called alternating direction method of multipliers.


\label{lastpage}

\end{document}